\documentclass[twocolumn,showpacs,aps,floatfix,prd,nofootinbib]{revtex4}
\usepackage{graphicx}
\usepackage{dcolumn}
\usepackage{array, longtable}
\usepackage{epsfig}
\usepackage{amsmath}
\usepackage{colordvi}
\usepackage{hhline}

\newcommand{\BaBarYear}{2004}
\newcommand{\BaBarNumber}{723}
\newcommand{\SLACPubNumber}{11026}

 \newcommand{\BaBarType}      {PUB}  

\input pubboard/babarsym

\long\def\inst#1{\par\nobreak\kern 4pt\nobreak
    {\it #1}\par\vskip 10pt plus 3pt minus 3pt}

\begin{document}


\begin{flushleft}
\babar-\BaBarType-\BaBarYear/\BaBarNumber \\
SLAC-PUB-\SLACPubNumber
\end{flushleft}


\title{\large \bf
\boldmath
The $\epem\to\pi^+\pi^-\pi^+\pi^-$,                 
$K^+K^-\pi^+\pi^-$, and $K^+K^- K^+K^- $
Cross Sections at Center-of-Mass Energies 0.5--4.5~\gev Measured with
Initial-State Radiation 
} 

%
\author{B.~Aubert}
\author{R.~Barate}
\author{D.~Boutigny}
\author{F.~Couderc}
\author{Y.~Karyotakis}
\author{J.~P.~Lees}
\author{V.~Poireau}
\author{V.~Tisserand}
\author{A.~Zghiche}
\affiliation{Laboratoire de Physique des Particules, F-74941 Annecy-le-Vieux, France }
\author{E.~Grauges-Pous}
\affiliation{IFAE, Universitat Autonoma de Barcelona, E-08193 Bellaterra, Barcelona, Spain }
% ...
\collaboration{The \babar\ Collaboration}
\noaffiliation

\date{\today}

\begin{abstract}
We study the process $e^+e^-\to\pi^+\pi^-\pi^+\pi^-\gamma$, with a hard
photon radiated from the initial state.  About 60,000 fully
reconstructed events have been selected from 89~\invfb of \babar\ data. The
invariant mass of the hadronic final state defines the effective \epem
center-of-mass energy, so that these
data can be compared with 
the corresponding direct \epem measurements. From the $4\pi$-mass spectrum, the
cross section for the process $\epem\to\pi^+\pi^-\pi^+\pi^-$ is measured
for center-of-mass energies from 0.6 to 4.5~\gev. The uncertainty in the cross
section measurement is typically 5\%. 
We also measure the cross sections for the final
states $K^+ K^- \pi^+\pi^-$ and $K^+ K^- K^+ K^-$. 
We observe the $J/\psi$ in all three final states and measure the corresponding branching fractions.  
We search for $X(3872)$ in $J/\psi (\to\mumu) \pipi$ and obtain an upper limit on the 
product of the \epem width of the $X(3872)$ and the branching fraction for $X(3872) \to J/\psi\pipi$.
\end{abstract}


\centerline{Submitted to Physical Review D}
\maketitle

\setcounter{footnote}{0}

\section{Introduction}
\label{sec:Introduction}

The idea of utilizing initial-state radiation (ISR) from a high-mass state
to explore electron-positron processes at all energies below that state was
outlined in Ref.~\cite{baier}.  The possibility of exploiting such processes
in high luminosity $\phi$- and $B$-factories was discussed in
Refs.~\cite{arbus, kuehn, ivanch} and motivates the study described in this
paper.  This is of particular interest because of the small discrepancy
between the measured muon $g-2$ value and that predicted by the Standard
Model~\cite{dehz}, where hadronic loop contributions are obtained from \epem
experiments at low center-of-mass (c.m.\@) energies. The study of ISR events
at $B$-factories provides independent and contiguous measurements of
hadronic cross sections in this energy region and also contributes to the
investigation of low-mass resonance spectroscopy.
                                
The ISR cross section for a particular hadronic final state $f$ (excluding
the radiated photon) is related to the 
corresponding \epem cross section $\sigma_f(s)$ by:
\begin{equation}
\frac{d\sigma_f(s,x)}{dx} = W(s,x)\cdot \sigma_f(s(1-x))\ ,
\end{equation}
where $x=2E_{\gamma}/\sqrt{s}$; $E_{\gamma}$ is the
energy of the ISR photon in the nominal \epem c.m.\@ frame; $\sqrt{s}$ is 
the nominal \epem c.m.\@ energy; and $\sqrt{s(1-x)}$ is the effective c.m.\@ energy
at which the final state $f$ is produced.
The function
\begin{equation}
W(s,x) = \beta\cdot \left[ (1+\delta)\cdot 
                           x^{(\beta-1)}-1+\frac{x}{2} \right]
\end{equation}
(see for example Ref.~\cite{ivanch}) describes the energy spectrum of the ISR 
photons, where $\beta =
2\alpha/\pi\cdot (2\ln (\sqrt{s}/m_e)-1)$ and $\delta$
takes into account vertex and self-energy corrections.  At the
$\Upsilon(4S)$ energy, 10.58~\gev, $\beta = 0.088$ and $\delta = 0.067$.  ISR
photons are produced at all angles. For
the present study it is required that the hard ISR photon be detected in the
electromagnetic calorimeter (EMC) of the \babar\ detector.  
Our acceptance for such photons is 10--15\%~\cite{ivanch}.

Events corresponding to $\epem\to\mumu\gamma$ provide the ISR-luminosity
normalization for the hadronic cross section measurements. For a hadronic
final state $f$, the Born cross section at center-of-mass energy
squared $s'$, $\sigma_f(s')$, is obtained by relating the observed number
of events, $dN_{f\gamma}$, in an interval $ds'$ centered at $s'$, to the
corresponding number of radiative di-muon events, $dN_{\mu\mu\gamma}$, by
means of
\begin{equation}
\sigma_f(s') = 
\frac{dN_{f\gamma}\cdot\epsilon_{\mu\mu}\cdot(1+\delta_{\rm FSR}^{\mu\mu})}
{dN_{\mu\mu\gamma}\cdot\epsilon_f\cdot(1+\delta_{\rm FSR}^{f})}
\cdot \sigma_{\epem\to\mumu}(s')\ .
\end{equation}
Here $s' \equiv s(1-x)$, $\epsilon_{\mu\mu}$ and $\epsilon_f$ are detection
efficiencies, and $(1+\delta_{\rm FSR}^{\mu\mu})$ and $(1+\delta_{\rm FSR}^{f})$ 
are 
corrections for the possibility that the detected hard photon may be the 
result of final-state radiation (FSR). This correction is important for
di-muon events, 
but is negligible for most hadronic final states. The Born cross section 
$\sigma_{\epem\to\mumu}(s')$ is used. The radiative corrections to the 
initial state, the acceptance for the ISR photon, and the virtual photon 
properties are the same for $\mumu$ and $f$, and cancel in the ratio.

An important advantage  of ISR data is that the entire
range of effective c.m.\@ energies is scanned in one experiment. 
This
avoids the relative normalization uncertainties that inevitably arise when
data from different experiments, or from different machine settings, are
combined.
                             
A disadvantage of the ISR measurement is that the mass resolution is
much poorer than can be obtained in a direct annihilation.
The resolution and absolute energy scale can be monitored
directly by the width and mass of the $J/\psi$ resonance produced in
the reaction $\epem \to J/\psi\gamma$. By using a kinematic fit to this
reaction, we find the resolution to be about 8~\mevcc for 
decays of $J/\psi$ in the $\mumu$ mode~\cite{Druzhinin1}.

Preliminary studies of $e^+e^-\to\mumu\gamma$ and some multihadron
ISR processes have been performed  
with \babar\ data~\cite{Sol, Druzhinin1,isr3pi}. These demonstrated good 
detector efficiency and particle identification capability for events of
this kind.

This paper reports analyses of the $\pi^+\pi^-\pi^+\pi^-$,
$K^+K^-\pi^+\pi^-$ and $K^+K^-K^+K^-$ final states produced in
conjunction with a hard photon, assumed to result from ISR.
While \babar\ data are available at effective c.m.\@ energies up to 10.58 \gev, 
the present analysis is restricted to 
energies below 4.5 \gev because of backgrounds from $\Upsilon(4S)$ decays.
A clear $J/\psi$ signal is observed for each of these hadronic states
and the corresponding $J/\psi$ branching fractions are measured.
A search for  the $X(3872)\gamma$ process  with the $X(3872)$ decay to
$J/\psi\pi^+\pi^-$ is also carried out.

\section{\boldmath The \babar\ detector and dataset}
\label{sec:babar}

The data used in this analysis were collected with the \babar\ detector at
the \pep2\ asymmetric \epem\ storage ring. The total integrated luminosity
used is 89~\invfb, which includes data collected at the $\Upsilon(4S)$
resonance mass (80~\invfb), and 
at c.m. energy 40 MeV lower (9~\invfb).

The \babar\ detector is described elsewhere~\cite{babar}. Final states with
four charged particles are reconstructed in the \babar\ tracking system,
which comprises the Silicon Vertex Tracker (SVT) and the drift chamber (DCH).
Separation of pions and kaons is accomplished by means of the Detector of
Internally Reflected Cherenkov Light (DIRC) and energy-loss measurements in
the SVT and DCH. The hard photon is detected in the EMC.  Muon
identification is provided by the Instrumented Flux Return (IFR) and this 
information is used to select the $\mumu\gamma$ final state and in  
the $X(3872)$ search.

The initial selection of candidate events requires that a high-energy photon
in the event with $E^\gamma_{\rm c.m.} > 3~\gev$ be found recoiling against
four good quality charged tracks with zero net charge. 
Events having such a high-energy photon 
together with an odd number ($\geq 3$) of good charged tracks are also
selected, for the purpose of making estimates of the tracking efficiency.
Each charged track is
required to originate close to the interaction region, to have transverse
momentum greater than 0.1~\gevc and to have a polar angle in the laboratory
frame with respect to the collision axis in the range from 0.4 to 2.45
radians. These selections guarantee the quality of the charged tracks in the DCH. 
Events with electrons and positrons are removed on the basis of associated
EMC-energy deposition and energy-loss (\dedx) information from the
DCH. 

In order to study the detector acceptance and efficiency, we developed a
special package of simulation programs for radiative processes.
The simulation of the
$\pi^+\pi^-\pi^+\pi^-\gamma$ final state is based on the generator developed 
by Kuehn  and Czyz~\cite{kuehn2}. The model assumes  $a_{1}(1260)\pi$ 
dominance~\cite{4pi_cmd, Bondar}, so that many of the events contain
a pair of 
pions from a $\rho$ meson due to the decay $a_{1}(1260)\to\rho^0\pi$. No 
corresponding generator exists for the $K^+ K^- \pi^+\pi^-$ and $K^+ K^- 
K^+ K^-$ final states, so these reactions were simulated according to      
phase space. 

Multiple soft-photon emission from the initial-state charged particles is
implemented with the structure-function technique~\cite{kuraev, strfun},
while extra photon radiation from the final-state particles is simulated by
means of the PHOTOS package~\cite{PHOTOS}.  The accuracy of the radiative
corrections is about 1\%.

A sample of about 400k events were generated with these tools and
passed through the detector response simulation \cite{GEANT4}. These
events were then reconstructed through
the same software chain as the experimental data. Variations in detector
and background conditions were taken into account.

For purposes of background estimation, a large sample of events from the
main ISR processes ($2\pi\gamma$, $3\pi\gamma$ ... $6\pi\gamma$, 
$2K\pi\gamma$ ...)
was simulated.  This sample exceeded the expected number of
events in the dataset by a factor of about three.  In addition, the 
expected numbers of $\epem\to q \qbar$ $(q = u, d, s, c)$ events  were generated via 
JETSET~\cite{jetset} and $\epem\to\tau^+\tau^-$ via KORALB~\cite{koralb}
in order to estimate non-ISR-type
background contributions. The cross sections for the above processes are known with
about 10\% accuracy or better, which is sufficient for the background contribution study.

\section{\boldmath The kinematic fit procedure}
\label{sec:Analysis}
The initial sample of candidate events is subjected to a constrained
kinematic fit in conjunction with charged-particle identification
to extract events corresponding to the final states of
interest.

For each particular four-charged-particle candidate, and for each possible
combination of particle types (i.e. $4\pi$, $2K2\pi$ or $4K$), a
one-constraint kinematic fit is performed without using information from the
detected photon candidate.  Because of the excellent resolution
of the DCH, the three-momentum vector of the photon is better determined
through momentum conservation than through measurement in the EMC. 
As a consequence, the calibration accuracy of the EMC and its alignment with 
respect to the DCH do not contribute to the systematic uncertainties.  
The initial \epem
and final-state charged-particle four-momenta and their covariance matrices
are taken into account.  The momentum vector of the photon reconstructed by
the fit in the laboratory frame is required to have polar angle $\theta^{\rm
fit}_{\gamma}$ in the range from 0.35 to 2.4~radians and to match the
measured polar angle $\theta^{\rm meas}_{\gamma}$ of a candidate photon in
the EMC within 50 mrad. The corresponding azimuthal angles, $\phi^{\rm
fit}_{\gamma}$ and $\phi^{\rm meas}_{\gamma}$, are also required to agree to
this same tolerance. These angular criteria reduce the background by a factor
of about two with no noticeable loss of signal. Finally, the polar angle
$\theta^{\rm fit}_{\rm ch}$ of each charged track after the fit has to
satisfy $0.45<\theta^{\rm fit}_{\rm ch}<2.4$~radians in order to fall within
the acceptance of the DIRC, which provides about 80\% kaon identification
efficiency.

The fit for the four-pion final-state hypothesis is
retained for every event. If only one track is identified as a kaon,
or if two oppositely-charged kaons are identified, the
$K^+K^-\pi^+\pi^-$ fit is also retained. Finally, if two,
three or four kaons are identified, the four-kaon fit is
applied.

For events with only three charged tracks recoiling against a
candidate photon, the measured four-momentum vector of the photon and
its covariance matrix are used in a one-constraint kinematic fit
which assumes, as appropriate, that only a charged pion or kaon is
undetected. These events are used in the efficiency studies described 
below. 

\begin{figure}[t]
\includegraphics[width=0.9\linewidth]{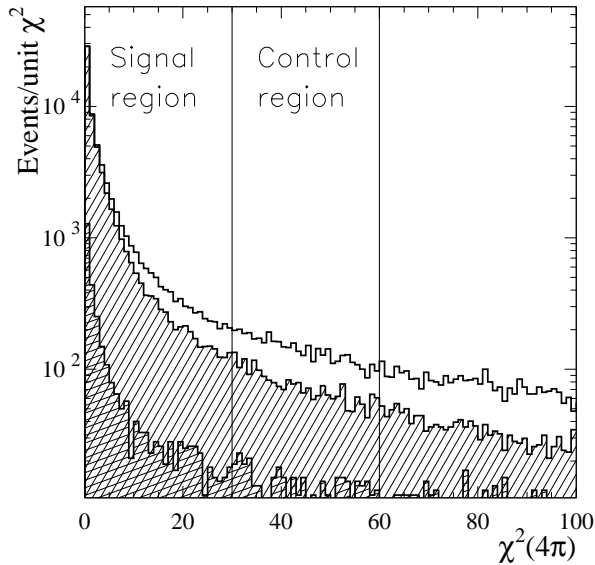}
\vspace{-0.3cm}
\caption{
The one-constraint \chisq distributions for data (upper histogram) and
       MC simulation (shaded histogram) four-charged-track events
       fitted to the four-pion hypothesis. The cross-hatched histogram
       is the estimated background contribution from non-ISR events
       obtained from JETSET. The signal and control regions are indicated.
}
\label{4pi_chi2_all}
\end{figure}
\begin{figure}[tbh]
\includegraphics[width=0.9\linewidth]{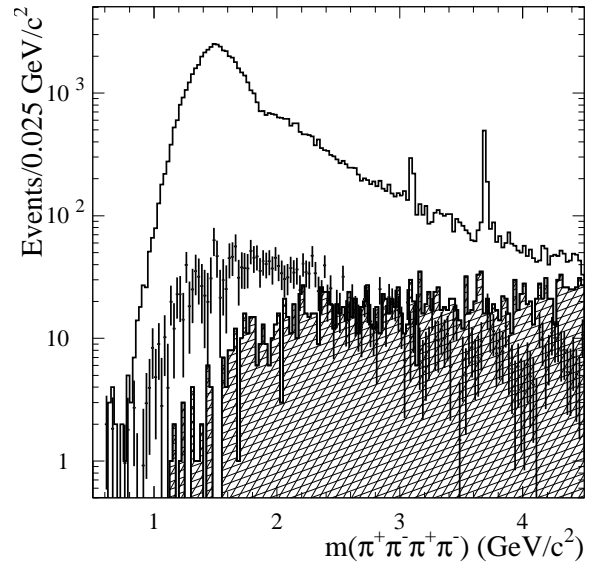}
\vspace{-0.3cm}
\caption{
The four-pion invariant mass distribution for the signal region of 
Fig.~\ref{4pi_chi2_all}. The points indicate the background estimated from
the difference between data and MC events for the control region of
Fig.~\ref{4pi_chi2_all}, normalized to the difference between data and MC
events in
the signal region of Fig.~\ref{4pi_chi2_all}. The cross-hatched histogram
corresponds to the non-ISR background of Fig.~\ref{4pi_chi2_all}.
}
\label{4pi_babar}
\end{figure}

\section{The {\boldmath $\pi^+\pi^-\pi^+\pi^-$} final state}
\subsection{Additional selection criteria}

The results of the one-constraint fit to the four charged-track candidates    
are used to make the final selection of the four-pion sample. 
We require $\chi_{4\pi}^2 < 30$ for the four-pion hypothesis, 
and that any accompanying fit to the 
$2K2\pi$ hypothesis have $\chi_{2K2\pi}^2 >10$. We estimate that 
 these requirements  reduce the contamination of the    
$4\pi$ sample by $2K2\pi$ events to about 1\% at the cost of about 2.4\% of 
the signal events. 

The one-constraint-fit \chisq distribution for the four-pion candidates is
shown as the upper histogram of Fig.~\ref{4pi_chi2_all}, while the shaded
region is for the corresponding MC-simulated pure $4\pi\gamma$ events.
The experimental distribution has a contribution from the background processes
but the MC-simulated distribution is also much broader than the usual
one-constraint \chisq distribution. This is due to multiple-soft-photon
emission in the initial state and radiation from the final-state charged
particles, neither of which is included in the constrained fit but which
exist in data and MC simulation.  To illustrate the difference of
distributions on Fig.~\ref{4pi_chi2_all}, the MC-simulated \chisq distribution is
normalized to the data in the region $\chi^2<1$ where contamination of the
background events and multiple soft ISR and FSR is lowest.

The cross-hatched histogram in
Fig.~\ref{4pi_chi2_all} represents the non-ISR
background contribution obtained from the JETSET simulation of
quark-antiquark production and hadronization and does not exceed 3\%.  

The region $30<\chi_{4\pi}^2<60$ is chosen as a control region for the
estimation of background from other ISR and non-ISR multi-hadron reactions. The
procedure followed is described in the next section.

The signal region of Fig.~\ref{4pi_chi2_all} contains 67,063 data             
and 71,210 MC events, while for the control region the corresponding           
numbers are 4,887 and 2,820 respectively.                                  
\subsection{Background estimation}
\label{sec:background}

MC simulation of the $\tau^+\tau^-$ final state and ISR production of
multi-hadron final states other than $\pi^+\pi^-\pi^+\pi^-$ shows that such
states would yield a background in the selected four-pion sample that would
exhibit a relatively flat contribution to the $\chi_{4\pi}^2$ distribution.
We subtract the shaded
histograms of Fig.~\ref{4pi_chi2_all} from the plain one and the resulting histogram 
is well described by MC simulation of background processes.
The background contribution to any distribution other
than \chisq is estimated as the difference between the distributions in the
relevant quantity for data and MC events from the control region of
Fig.~\ref{4pi_chi2_all}, normalized to the difference between the number of
data and MC events in the signal region.

For example, Fig.~\ref{4pi_babar} shows the four-pion invariant mass
distribution up to 4.5~\gevcc for the signal region of
Fig.~\ref{4pi_chi2_all}. The points with error bars show the ISR background
contribution obtained in the manner described from the control region of
Fig.~\ref{4pi_chi2_all}.  
The cross-hatched histograms in
Fig.~\ref{4pi_chi2_all} and Fig.~\ref{4pi_babar} represent the non-ISR
background contribution obtained from the JETSET MC simulation.
The \chisq distribution for the non-ISR
events is not flat, and 
we estimate their relative contribution to the signal region
from the production cross-section and the integrated luminosity.
Both backgrounds are small at low mass, but the non-ISR
background accounts for almost half of the observed data at approximately
4\gevcc.  The data show a strong peak around 1.5~\gevcc followed by a
shoulder near 1.9~\gevcc. Narrow signals are apparent at the $J/\psi$ and
the $\psi(2S)$ masses, although the latter is due to $\psi(2S)\to\pipi
J/\psi$ $J/\psi\to\mumu$ with the muons being treated as pions.

Accounting for uncertainties in cross sections for background processes and 
statistical fluctuations in the number of simulated events,
we estimate that this procedure for background subtraction results
in a systematic uncertainty of less than 1\% in the number of signal
events in the 1--3~\gevcc region of four-pion mass, but that it            
increases to 3--5\% in the region above 3~\gevcc and to roughly 10\%
in the region below 1~\gevcc. 

By selecting a ``background-free'' $4\pi\gamma$ sample with only four charged
tracks and only one photon (about 10\% of events) we can compare \chisq
distributions for data and MC events up to \chisq=1000.  We estimate that
for a $\chi_{4\pi}^2<30$ selection the net signal size should be increased by
$(3\pm2)\%$ to allow for a slight shape difference between the MC and experimental
\chisq distributions.
\begin{figure}[tbh]
\includegraphics[width=0.9\linewidth]{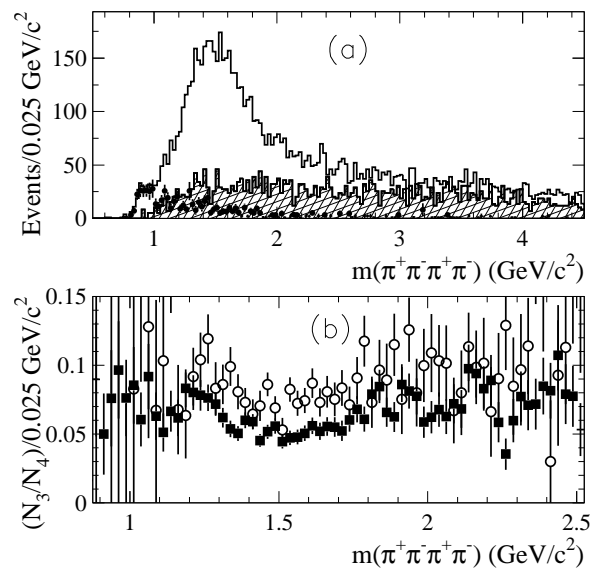}
\vspace{-0.3cm}
\caption{
(a) The four-pion invariant mass distribution obtained from
 fits to three-charged-track events in which an observed photon is
 used to constrain the unobserved fourth pion; ISR and non-ISR background
 contributions are indicated by the points with error bars and
 the cross-hatched histogram, respectively. (b) The four-pion
 mass dependence of the ratio between the three- and four-charged-track
 distributions for data (open circles) and MC simulation (solid squares).
}
\label{m4pi_3tr}
\end{figure}
\begin{figure}[tbh]
\includegraphics[width=0.9\linewidth]{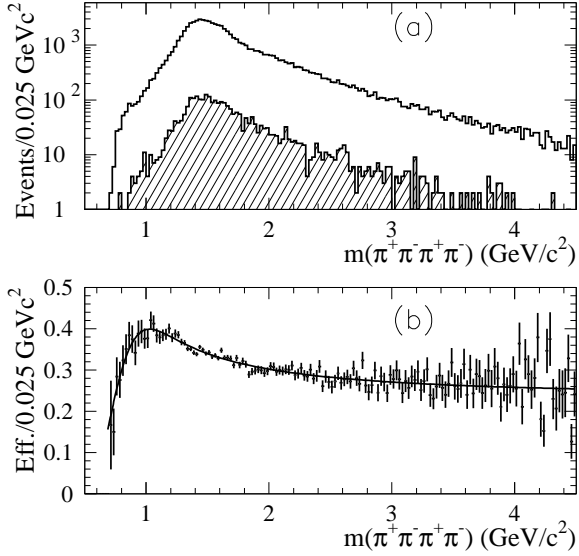}
\vspace{-0.3cm}
\caption{
(a) The four-pion mass distribution from MC simulation for the
  signal and control (shaded) regions of Fig.~\ref{4pi_chi2_all}. (b) The mass
  dependence of the net reconstruction and selection efficiency
  obtained from simulation.
}
\label{mc_acc}
\end{figure} 
\subsection{Tracking efficiency}
\label{sec:tracking}
We measure the track-finding efficiency
with events that have three charged-particle tracks and a hard
photon. These  events are subjected to a
one-constraint fit (as described in Sec.~\ref{sec:Analysis} above), which
yields the
three-momentum vector of the missing charged pion in the laboratory frame
assuming this is the only undetected track. If the chi-squared of the fit is 
less than 30 and
this vector lies within the
acceptance of the DCH, the event is included in the data sample.

The four-pion mass distribution obtained in this way for three-charged-track
events is shown in Fig.~\ref{m4pi_3tr}(a), together with the ISR (points
with errors) and non-ISR (cross-hatched histogram) background estimated
as described above. The behavior is similar to that
observed in Fig.~\ref{4pi_babar}. This is exhibited explicitly in
Fig.~\ref{m4pi_3tr}(b), where the ratio of the three- to four-charged track
pion mass distributions (after background subtraction) is shown as a
function of four-pion invariant mass for data (open points). When the 
MC-simulated data are treated in the same way, the solid points of
Fig.~\ref{m4pi_3tr}(b) are obtained. The same absence of mass dependence is
observed for data and MC events, but the MC simulation yields a smaller
fraction of three-track events than observed for the data.  This difference
is $(3.0\pm0.3\pm2.0)\%$, where the systematic error is
estimated from the slight difference in the mass dependence 
seen in Fig.~\ref{m4pi_3tr}(b), from the
uncertainties in background subtraction and from a slight difference in angular 
dependence.
This uncertainty
increases to about 10\% for the mass region below 1~\gevcc. The
systematic difference is used to correct the observed signal size for the
difference in net track-finding efficiency between that obtained from MC
simulation and that observed in the experiment.
\subsection{Detection efficiency from simulation}
\begin{figure}[tbh]
\includegraphics[width=1.05\linewidth]{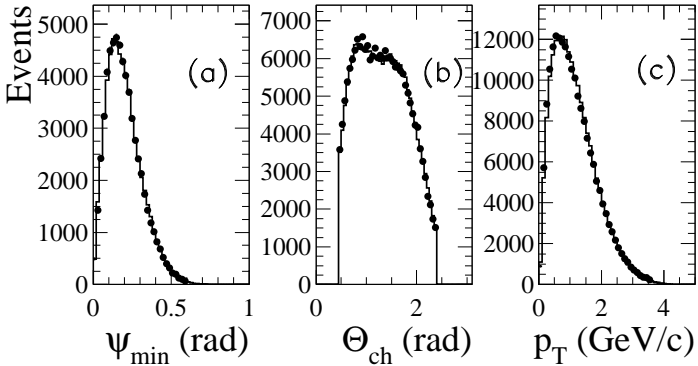}
\vspace{0.1cm}
\caption{
(a) The distribution in track-pair opening angle for the
minimum of the six values possible for each event; (b) the
        distribution in polar angle, and (c) the transverse momentum
        distribution for all pions from all events. All quantities
        are in the laboratory frame; the points are for data and the
        histograms are obtained from MC simulation.
}
\label{mc_distr}
\end{figure}
\begin{figure}[tbh]
\includegraphics[width=1.05\linewidth]{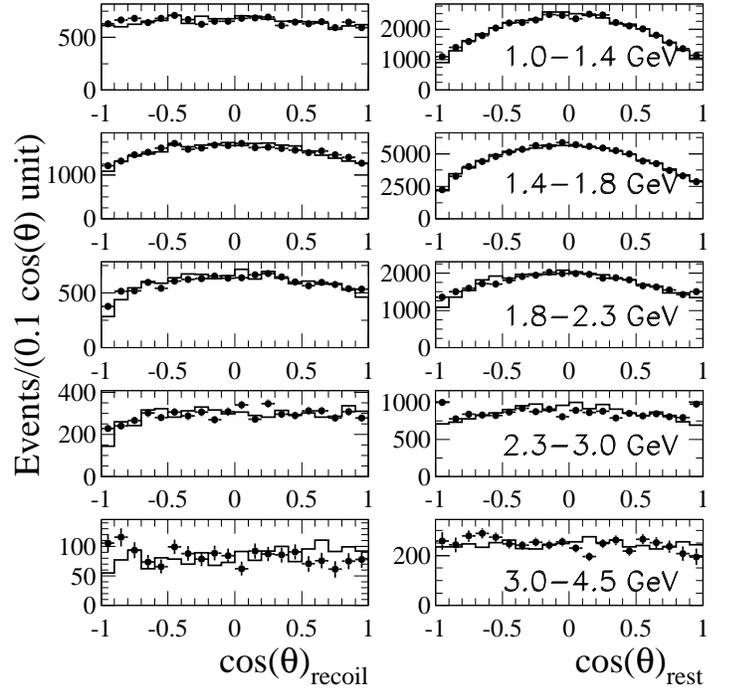}
\vspace{-0.3cm}
\caption{
The angular distribution of the lowest-momentum pion (left)
and of the three most energetic pions (right) in the four-pion rest
frame  with respect to the direction of the four-pion system
       in the laboratory frame for the five regions of four-pion mass
       indicated. The points are data, and the histograms are
       MC simulation. 
}
\label{angle_slices}
\end{figure}
The selection procedures applied to the data are also applied to the         
MC-simulated event sample. The resulting four-pion invariant-mass
distribution is shown in Fig.~\ref{mc_acc}(a) for the signal and
control (shaded histogram) regions. There is good qualitative
agreement with the mass distribution of Fig.~\ref{4pi_babar}, except
that no attempt was made to simulate the $J/\psi$ and $\psi(2S)$
signals observed in the data. The mass dependence of the detection
efficiency is obtained by dividing the number of reconstructed MC
events in each 25~\mevcc mass interval by the number generated in          
this same interval. The result is shown in Fig.~\ref{mc_acc}(b); the 
curve is obtained from a polynomial fit to the distribution. The              
efficiency increases from 20\% at about 0.8~\gevcc to a maximum 
of 40\% near 1~\gevcc, and thereafter falls off gradually 
with increasing mass to about 26\% at 4.5~\gevcc. This efficiency 
estimate takes into account the geometrical acceptance of the detector 
for the final-state photon and the charged pions, the inefficiency of 
the  several detector subsystems and event-loss due to additional
soft-photon  emission from the initial and final states.

As mentioned in Sec.~\ref{sec:babar}, the model used in the MC simulation 
assumes
that the four-pion final state results predominantly from the
$a_{1}(1260)\pi$ quasi-two-body production process~\cite{kuehn2}. A
contribution from  $f_{0}(1370)\rho(770)$ is incorporated also. In
general, this model describes well the
distributions in many of the kinematic variables characterizing the
four-pion final state. Some examples are shown in Figs.~\ref{mc_distr}
and~\ref{angle_slices}, in which the points with error bars represent
data while the histograms are obtained from MC simulation; for
both figures, the $J/\psi$ and $\psi(2S)$ regions have been
excluded. Figure~\ref{mc_distr}(a) shows the distribution in
$\psi_{\rm min}$, the minimum charged-pion-pair opening angle for each
event, while Fig.~\ref{mc_distr}(b) and Fig.~\ref{mc_distr}(c)
represent the distribution in polar angle, $\theta_{\rm ch}$, and
transverse momentum, $p_T$, respectively, for all final-state
pions. All quantities are calculated in the laboratory frame, and the        
overall agreement between MC simulation and data is very
good. Figure~\ref{angle_slices} 
compares the distributions in $\cos\theta$, where $\theta$ 
is the angle between a charged pion in the four-pion rest frame, and 
the direction of the four-pion system in the laboratory frame. The           
distributions are presented for the five regions of four-pion mass indicated
(mass increasing from top to bottom); the left column is for the 
lowest-momentum pion in the four-pion rest frame, and the right column sums         
the distributions for the others. Data and MC are in relatively good 
agreement up to about 2~\gevcc, but but above this value some small 
discrepancies appear. 

In the four-pion rest frame, the angular acceptance is rather
uniform. Changes in the $a_{1}(1260)$ and $f_{0}(1370)$ resonance
parameter values within the ranges of their uncertainties produce            
little effect. However, simulation without resonances using only
four-pion phase space
does produce discernible deviations from the observed angular
distributions, and changes the overall acceptance by about 2\%. This
value is taken as an estimate of systematic uncertainty in the acceptance
associated with the simulation model used. 

\begin{figure}[tbh]
\includegraphics[width=0.9\linewidth]{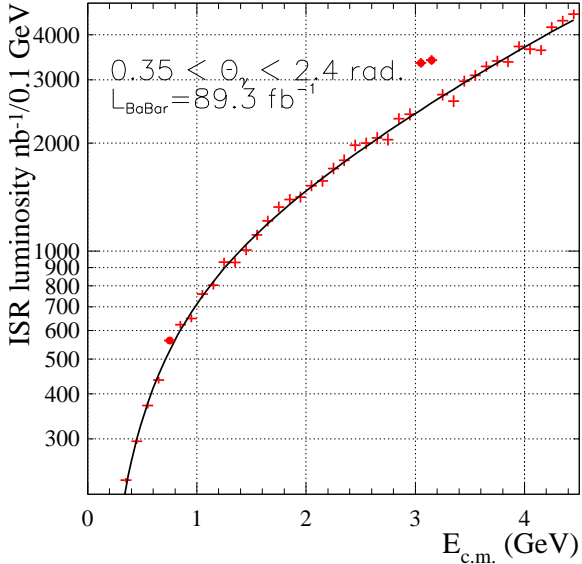}
\vspace{-0.3cm}
\caption{
The integrated ISR luminosity per 0.1~\gev in effective c.m.\@          
energy obtained from $\ep\ep\to\mu^+\mu^-\gamma$ events. The curve
       represents the fit used in the luminosity calculations. The
       point at the $\rho(770)$ mass position was excluded from the 
       fit because of pion misidentification feed-through from the 
       $\pi^+\pi^-\gamma$ final state. The points at the $J/\psi$ 
       mass position were also excluded because of ISR production of
       $J/\psi$ followed by decay to $\mu^+\mu^-$.
}
\label{isr_lumi}
\end{figure}
\begin{figure}[tbh]
\includegraphics[width=0.9\linewidth]{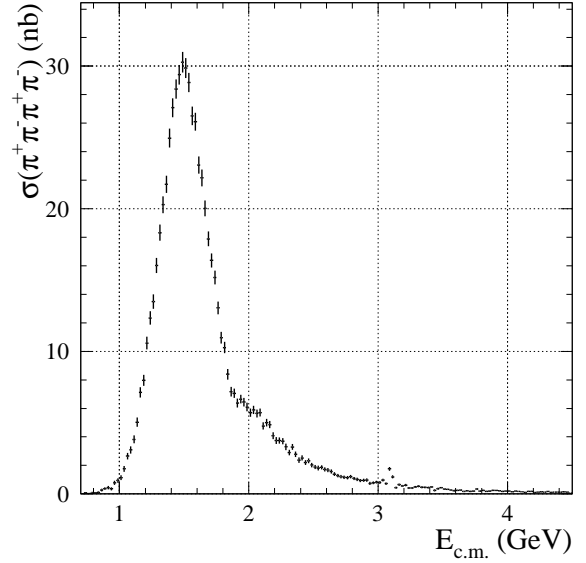}
\vspace{-0.3cm}
\caption{
The \epem c.m.\@ energy dependence of the $\pipi\pipi$ cross section
 measured with ISR data. The events due to $\psi(2S)\to\pipi J/\psi$
 with $J/\psi\to \mumu$ (see Fig.~\ref{4pi_babar}) have been removed.
Only statistical errors are shown.
}
\label{4pi_ee_babar}
\end{figure} 
\subsection{\boldmath Cross section for $\epem\to\pi^+\pi^-\pi^+\pi^-$}
Data from the reaction $\epem\to\mumu\gamma$ are used to convert the
invariant-mass distribution for an ISR-produced hadronic final state to        
the energy dependence of the corresponding \epem cross section.
The invariant mass of the muon pair $m_{\rm inv}^{\mu\mu}$ defines an
effective \epem c.m.\@ collision energy, $E_{\rm c.m.}$. The differential
luminosity, $d{\cal L}$, associated with the interval $dE_{\rm c.m.}$ centered at
effective collision energy $E_{\rm c.m.}$ is then obtained from
\begin{equation}
  d{\cal L}(E_{\rm c.m.})
  = \frac{dN_{\mu\mu\gamma}(E_{\rm c.m.})}
         {\epsilon_{\mu\mu}\cdot(1+\delta_{\rm FSR}^{\mu\mu})
          \cdot\sigma_{\mumu}(E_{\rm c.m.})\cdot(1+\delta_{\rm vac})}\ ,
\end{equation}
where $E_{\rm c.m.}=m_{\rm inv}^{\mu\mu}$; $dN_{\mu\mu\gamma}$ is the number of
muon pairs in the mass interval $dm_{\rm inv}^{\mu\mu}=dE_{\rm c.m.}$; 
$\epsilon_{\mu\mu}$ is the acceptance, corrected for muon identification and
soft-photon emission; $(1+\delta_{\rm FSR}^{\mu\mu})$ corrects for hard photon
emission from final-state muons;
$~\sigma_{\mumu}(E_{\rm c.m.})$  is the $\epem\to\mumu$ Born cross section at 
center-of-mass energy $E_{\rm c.m.}$; and $(1+\delta_{\rm vac})$ is the 
corresponding vacuum
polarization correction~\cite{EJ}.  For the $\mumu\gamma$ sample obtained from 
a \babar\ 
integrated luminosity of 89~\invfb, the dependence of the resulting
differential luminosity on $E_{\rm c.m.}$ is shown in Fig.~\ref{isr_lumi} in
units of [$\invnb/0.1~\gev$]. From a detailed study of the
$\epem\to\mumu\gamma$ detection and identification efficiency described in
detail in Ref.~\cite{Druzhinin1} and comparison of the observed invariant-mass
spectrum with theoretical calculations, we estimate the systematic uncertainty
associated with luminosity determination to be 3\%.

The four-pion \epem cross section can then be calculated from
\begin{equation}
  \sigma(\pipi\pipi)(E_{\rm c.m.})
  = \frac{dN_{4\pi\gamma}(E_{\rm c.m.}^\pi)}
         {d{\cal L}(E_{\rm c.m.})\cdot\epsilon_{4\pi}^{\rm corr}
          \cdot\epsilon_{4\pi}^{\rm MC}(E_{\rm c.m.}^\pi)}\ ,
\end{equation}
where $E_{\rm c.m.}^\mu \equiv m_{\rm inv}^{\mu\mu} \equiv E_{\rm c.m.}^\pi \equiv
m_{\rm inv}^{4\pi} \equiv E_{\rm c.m.}$ with $m_{\rm inv}^{4\pi}$ the invariant mass of
the four-charged-pion system; $dN_{4\pi\gamma}$ is the number of selected
four-pion events after background subtraction in the interval $dE_{\rm c.m.}$ and
$\epsilon_{4\pi}^{\rm MC}(E_{\rm c.m.})$ is the corresponding detection
efficiency obtained from the MC simulation. The factor
$\epsilon_{4\pi}^{\rm corr}$ takes into account the 
difference between the \chisq distributions for data and MC events,
and the tracking-efficiency discrepancies 
discussed in Sec.~\ref{sec:background} and Sec.~\ref{sec:tracking} respectively.

Since the four-pion cross section calculation involves the ratio of the
numbers of
observed $4\pi\gamma$ and $\mu\mu\gamma$ events, corrections related to 
multi-soft-photon emission in the initial state and to the detection 
efficiency of the ISR photon cancel, these being the same for both 
reactions. Also, since $d{\cal L}$ has been corrected for vacuum polarization 
and final-state soft-photon emission, the four-pion 
cross section measured in this way includes effects due to vacuum 
polarization and final-state soft-photon emission.

We studied the resolution in four-pion mass with MC simulation and found
that the r.m.s.\@ deviation varied from 6.2~\mevcc at mass 1.5~\gevcc to
7.5~\mevcc at 3~\gevcc. Since the cross section 
has no sharp peaks ($J/\psi$ region is discussed below) and the 
measurements are presented
in mass intervals of 25~\mevcc, the
resolution has negligible effect on the measured energy
dependence.
 
The energy dependence of the cross section for the reaction
$\epem\to\pi^+\pi^-\pi^+\pi^-$ after all corrections is shown in
Fig.~\ref{4pi_ee_babar}. It reaches a peak value of about 30~nb near
1.5~\gev, with a shoulder at 1.9--2.1~\gev, followed by a monotonic decrease toward
higher energies perturbed only by a small peak at the $J/\psi$ mass position.
The events due to $\psi(2S)\to\pipi J/\psi$ with $J/\psi\to \mumu$, seen in
Fig.~\ref{4pi_babar}, have been removed.  The luminosity, number of events,
and corrected cross section for each 25~\mev interval are presented in
Table~\ref{4pi_tab}.  For $g-2$ calculations, vacuum polarization
contributions should be excluded. Suitably modified cross section values are
presented in the last column of Table~\ref{4pi_tab}.

We checked the stability of the measured cross section by comparing the
results obtained under conditions of different DCH voltage, which
affects track-finding efficiency, and results obtained from data taken
at the $\Upsilon(4S)$ mass and below the \BB\ threshold. No significant
discrepancies were observed.

\begin{figure}[tbh]
\includegraphics[width=0.9\linewidth]{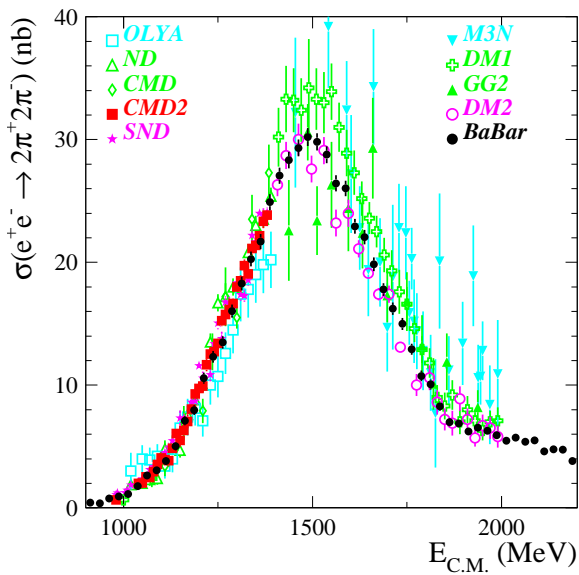}  
\vspace{-0.3cm}
\caption{
 The energy dependence of the $\epem\to\pipi\pipi$ cross section
obtained with \babar\ ISR data (black points) in comparison
with that from direct \epem production measurements.
Only statistical errors are shown.
}
\label{xs_4pi_all}
\end{figure}
\begin{figure}[tbh]
\includegraphics[width=0.9\linewidth]{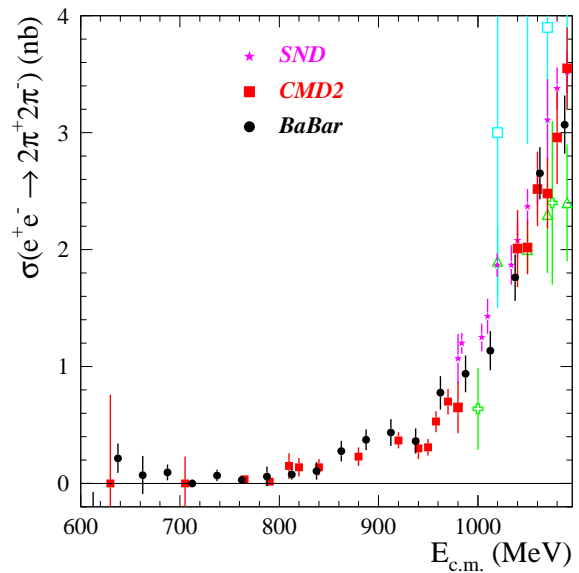}
\vspace{-0.3cm}
\caption{
A detailed view of the energy dependence of the
$\epem\to\pipi\pipi$
cross section near threshold; All
experimental points are indicated as in Fig.~\ref{xs_4pi_all}.
Only statistical errors are shown.
}
\label{xs_4pi_low}
\end{figure}
\subsection{\boldmath Summary of systematic studies}
\label{sec:Systematics}
The measured four-pion cross section values shown in Fig.~\ref{4pi_ee_babar} and 
summarized in Table~\ref{4pi_tab} include only statistical errors. The systematic 
errors discussed in previous sections are summarized in Table~\ref{error_tab}, along 
with the corrections that were applied to the measurements.
\begin{table*}[tbh]
\caption{
Summary of systematic errors for $\epem\to\pipi\pipi$ cross section
}
\label{error_tab}
\begin{ruledtabular}
\begin{tabular}{l c l} 
Source & Correction applied & Systematic error\\
\hline
Luminosity from $\mu\mu\gamma$ &  -  &  $3\%$ \\
                               &     &  $5\%$ for $m_{4\pi}<1.0~\gevcc$ \\
MC-data difference in $\chi^{2}<30$ requirement & $+3\%$ & $2\%$\\ 
Background subtraction & - &  $1\%$ \\
                       &   &  $10\%$ for $m_{4\pi}< 1.0~\gevcc$\\
                       &   &  $3\%$ for $m_{4\pi}>3.0~\gevcc$ \\
MC-data difference in track losses & $+3\%$ & $2\%$ \\
Radiative corrections accuracy & - & $1\%$ \\
Acceptance from MC (model-dependent) & - & $2\%$ for
$<3~\gev$ \\
                                       &  & $15\%$ for the rest\\
\hline
Total     &  $+6\%$   & $12\%$ for $m_{4\pi}<1.0~\gevcc$ \\
(assuming no correlations) &   & $5\%$ for $1.0<m_{4\pi}<3.0~\gevcc$\\
                          &   & $16\%$ for $m_{4\pi}>3.0~\gevcc$\\
\end{tabular}
\end{ruledtabular}
\end{table*}

The two systematic corrections applied to the measured cross sections
sum up to +6\% with half of this value taken as a systematic uncertainty. 
The systematic errors that cancel in the ratio to $\mu\mu\gamma$ are
the photon detection efficiency and the ISR soft-photon radiative
correction uncertainty.

\subsection{\boldmath Physics results}
\label{sec:Physics}
The four-charged-pion cross section measured by \babar\ can be compared with
existing \epem measurements only up to {2.0~\gev}---the maximum c.m.\@ energy
up to which measurements of this channel have been published.
Figures~\ref{xs_4pi_all} and \ref{xs_4pi_low} show the cross section in
comparison with all existing \epem data for c.m.\@ energies in the
0.7--2.2~\gev range.

The measured cross section is in good agreement with the precision data
taken at VEPP-2M by SND~\cite{4pi_snd} and CMD-2~\cite{4pi_cmd, 4pi_cmd_low}
in the energy range 0.7--1.4~\gev, as well as with data obtained at DCI by
DM2~\cite{4pi_dm2} in 1.4--2.0~\gev range. The systematic errors for \babar\ 
data are comparable to, or smaller than, those for these experiments.

\begin{figure}[tbh]
\includegraphics[width=1.0\linewidth]{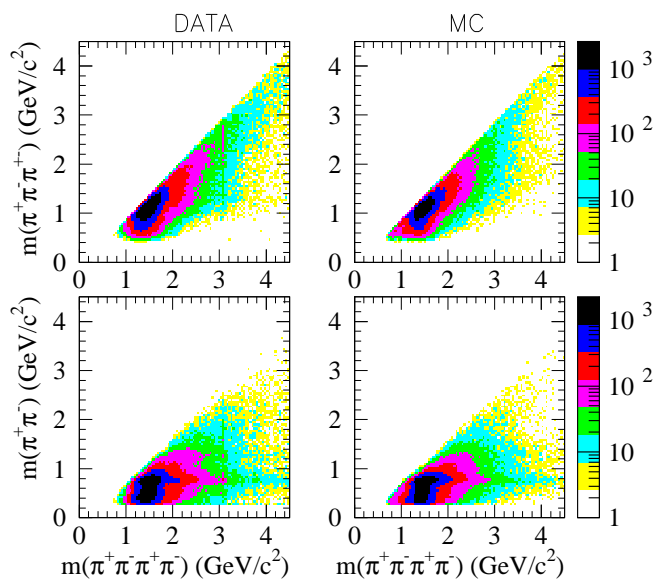}
\vspace{-0.3cm}
\caption{
The two-pion (bottom) and three-pion (top) vs. four-pion invariant mass
distributions for data (left) and simulation (right);
        signals corresponding to $J/\psi$ and $\psi(2S)$ production are
        present in the data, but are not included in the simulation.
        Otherwise, agreement between data and MC simulation is quite good.
}
\label{4pi_2-3pi}
\end{figure}

Different mass combinations were studied in data and MC events to       
search for any structures or states not included in the simulation.                
Figure~\ref{4pi_2-3pi} shows the scatter-plots of 3$\pi$- and 2$\pi$-mass
versus 4$\pi$-mass for data and MC events.  Good agreement is seen          
except for narrow regions around the $J/\psi$ and $\psi(2S)$ masses, which       
are not included in the simulation.                                           

\begin{figure*}[tb]
\includegraphics[width=0.64\linewidth]{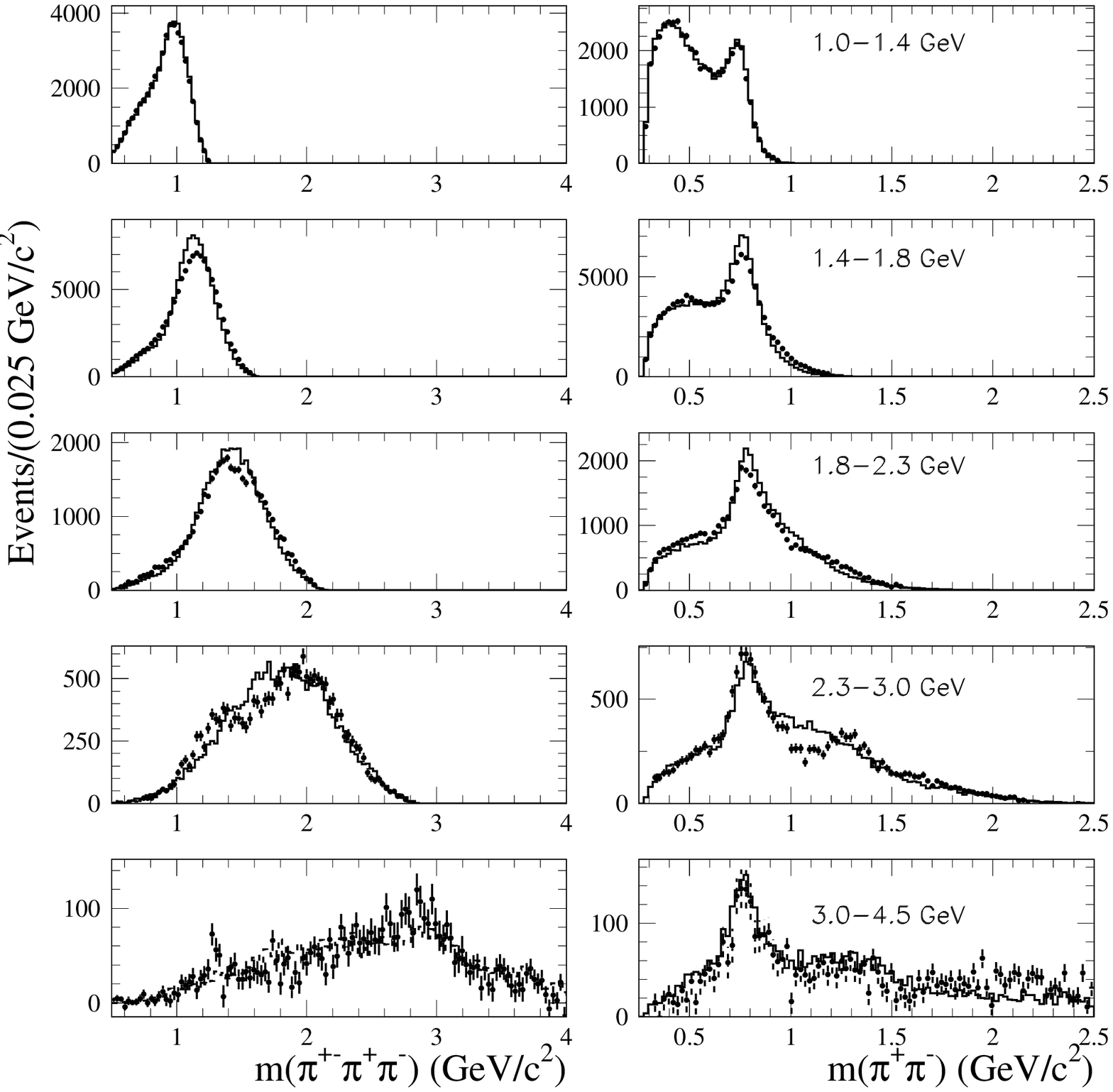}
\hfill
\includegraphics[width=0.35\linewidth]{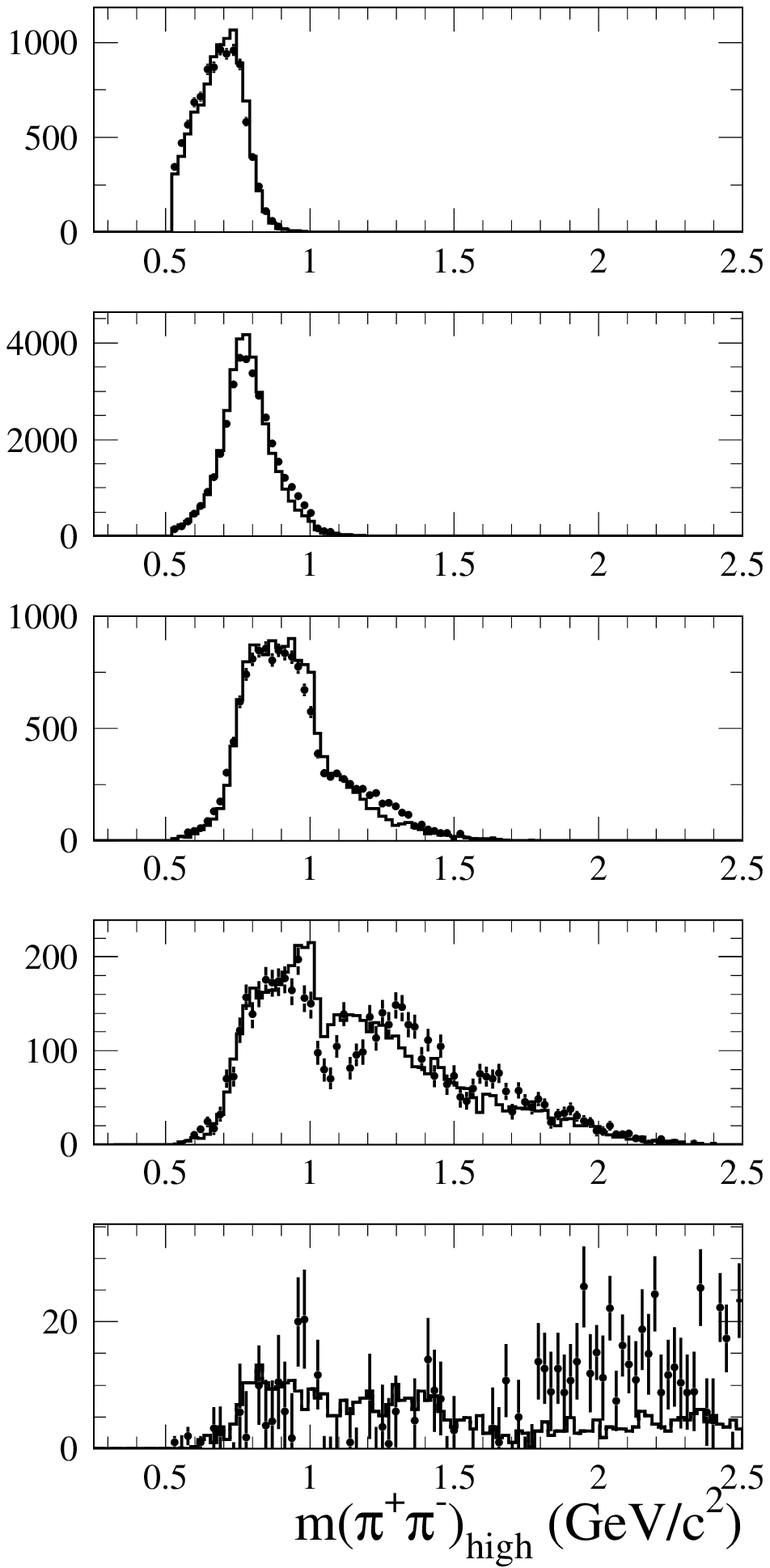}
\vspace{-0.3cm}
\caption{
        The three-pion (left) and $\pi^+\pi^-$ (center) mass
        distributions; the points represent
        data, the histograms simulation, and there are four entries
        per event. 
        Five regions of four-pion mass are indicated in the center
        plots.
        For events having a $\pi^+\pi^-$ mass combination in the 
        $\rho(770)$ region, the distribution of the highest mass other 
        $\pi^+\pi^-$ pair is shown in the plots to the right. 
}
\label{2pi_3pi_pro}
\end{figure*}

In order to make a more detailed study, five intervals of 4$\pi$-mass are
selected: (1) 1.0--1.4~\gevcc (for comparison with CMD-2), (2) 1.4--1.8~\gevcc
(peak in cross section, see Fig.~\ref{4pi_ee_babar}), (3) 1.8--2.3~\gevcc 
(shoulder), (4) 2.3--3.0~\gevcc,
and (5) 3.0--4.5~\gevcc (narrow regions around the $J/\psi$ and $\psi(2S)$ are
excluded). 
Figure~\ref{2pi_3pi_pro} shows the one-dimensional projections of 
Fig.~\ref{4pi_2-3pi} of the $2\pi$- and $3\pi$-mass 
for the above five regions for data and MC events. 
In these distributions we subtract the background using control samples 
in the \chisq
distributions from data and JETSET simulation, as described
above. 

There is excellent agreement between data and MC events  in the 1.0--1.4~\gevcc 
region where
$a_{1}(1260)$ production is severely limited by the available phase
space, and $f_{0}(1370)$ production is almost entirely
excluded. Discrepancies begin to appear in the higher mass regions,
and in particular a relatively narrow bump in the $\pipi$ combinations
at about 1.3~\gevcc for the 2.3--3.0~\gevcc region is not
reproduced by the simulation. 
The world averages reported by the Particle Data Group (PDG)~\cite{PDG}
for the $a_{1}(1260)$
mass and width are not well-determined; in the simulation we use
1.33~\gevcc  and 0.57~\gev respectively. These values were obtained from
a combined analysis of CLEO and CMD-2 data~\cite{Bondar}.

The 3$\pi$-mass data of Fig.~\ref{2pi_3pi_pro} (1.8-2.3~\gevcc) seem to favor 
a lower $a_{1}(1260)$ mass value than the 1.33~\gevcc used in simulating
the $a_{1}(1260)$.
In the simulation, the $f_{0}(1370)$ mass and
width were set to 1.3~\gevcc and 0.6~\gev respectively.  The 2$\pi$-mass data for
the 2.3--3.0~\gevcc region seem to indicate the need for a
significantly smaller width, although the peak around 1.3~\gevcc may also
be due to significantly higher production of $f_{2}(1270)$ than in the
present model.

In Fig.~\ref{2pi_3pi_pro}, the right-hand column of plots corresponds to the
highest di-pion mass value when one other di-pion mass is within 25~\mevcc of the
$\rho$ mass. The flat, broad band below 1~\gevcc is the reflection of the $\rho$
band in $a_{1}(1260)$ decay. The comparatively narrow peak around
1.3~\gevcc in the 2.3--3.0~\gevcc region indicates the need for
$f_{0}(1370)\rho$ in the simulation.

A full partial wave analysis  is required in order to arrive at a more
precise interpretation of the data. However, this requires a simultaneous
analysis of the $\pi^+\pi^-\pi^0\pi^0$ final state, which is beyond the scope of 
this study.
\begin{figure}[tbh]
\includegraphics[width=0.9\linewidth]{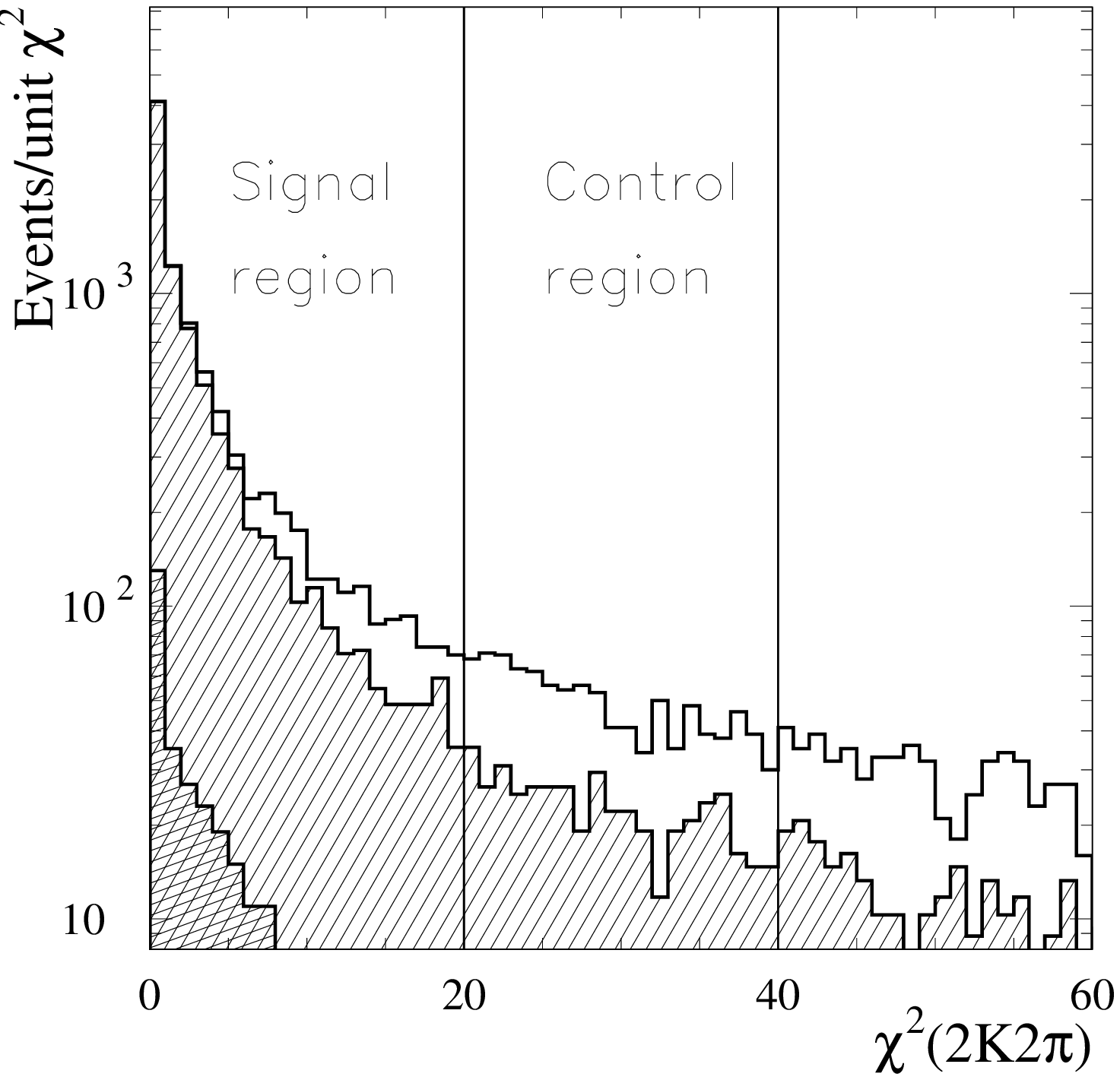}
\vspace{-0.3cm}
\caption{
       The one-constraint \chisq distributions for the
       four-charged-track data events and $K^+ K^-\pi^+\pi^-$ Monte 
       Carlo events (shaded histogram) fitted to the 
       $K^+ K^-\pi^+\pi^-$ hypothesis. At least one kaon must 
       be identified. The cross-hatched histogram is the 
       estimated background contribution from four-charged-pion 
       ISR events. The signal and control regions 
       are indicated.
}
\label{chi2_2k2pi}
\end{figure}
\begin{figure}[tbh]
\includegraphics[width=0.9\linewidth]{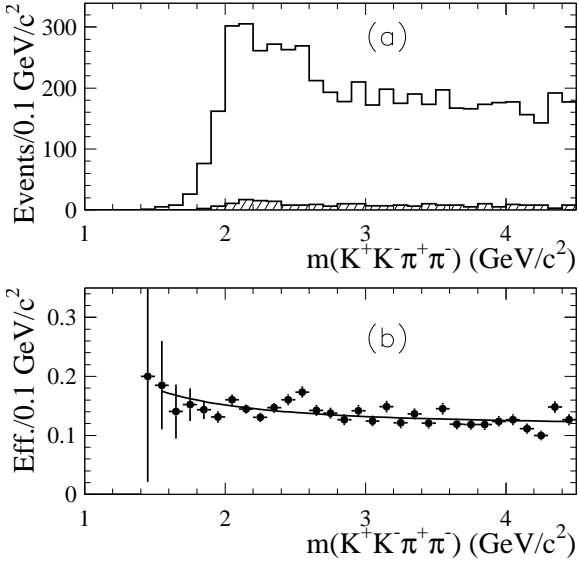}
\vspace{-0.3cm}
\caption{
(a) The $K^+ K^-\pi^+\pi^-$ mass distribution from simulation 
for the signal and control (shaded) regions of Fig.13; (b) the 
mass dependence of the net reconstruction and selection 
efficiency obtained from simulation.
}
\label{2k2pi_acc}
\end{figure} 
\section{\boldmath The $K^+K^-\pi^+\pi^-$ final state}
The constrained fit of the four-charged-track events to the hypothesis of
two oppositely charged kaons and two charged pions, where
at least one of the kaons has positive particle identification, allows
us to select this final state.
Figure~\ref{chi2_2k2pi} shows the \chisq
distributions for both data and simulation, where the simulation of the 
$K^+K^-\pi^+\pi^-$ reaction uses a point-like matrix 
element
with a cross section energy dependence close to that which we observe 
experimentally, and
all  radiative processes are
included. Also shown is the estimated contribution from the $4\pi$ final 
state
arising from particle misidentification, as obtained from simulation.

Figure~\ref{2k2pi_acc}(a) presents the simulated mass distribution for
the $2K2\pi$ events; the mass dependence of the
efficiency, calculated as a ratio of selected  
to generated $2K2\pi$ MC events, is shown in Fig.~\ref{2k2pi_acc}(b).
\begin{figure}[tbh]
\includegraphics[width=0.9\linewidth]{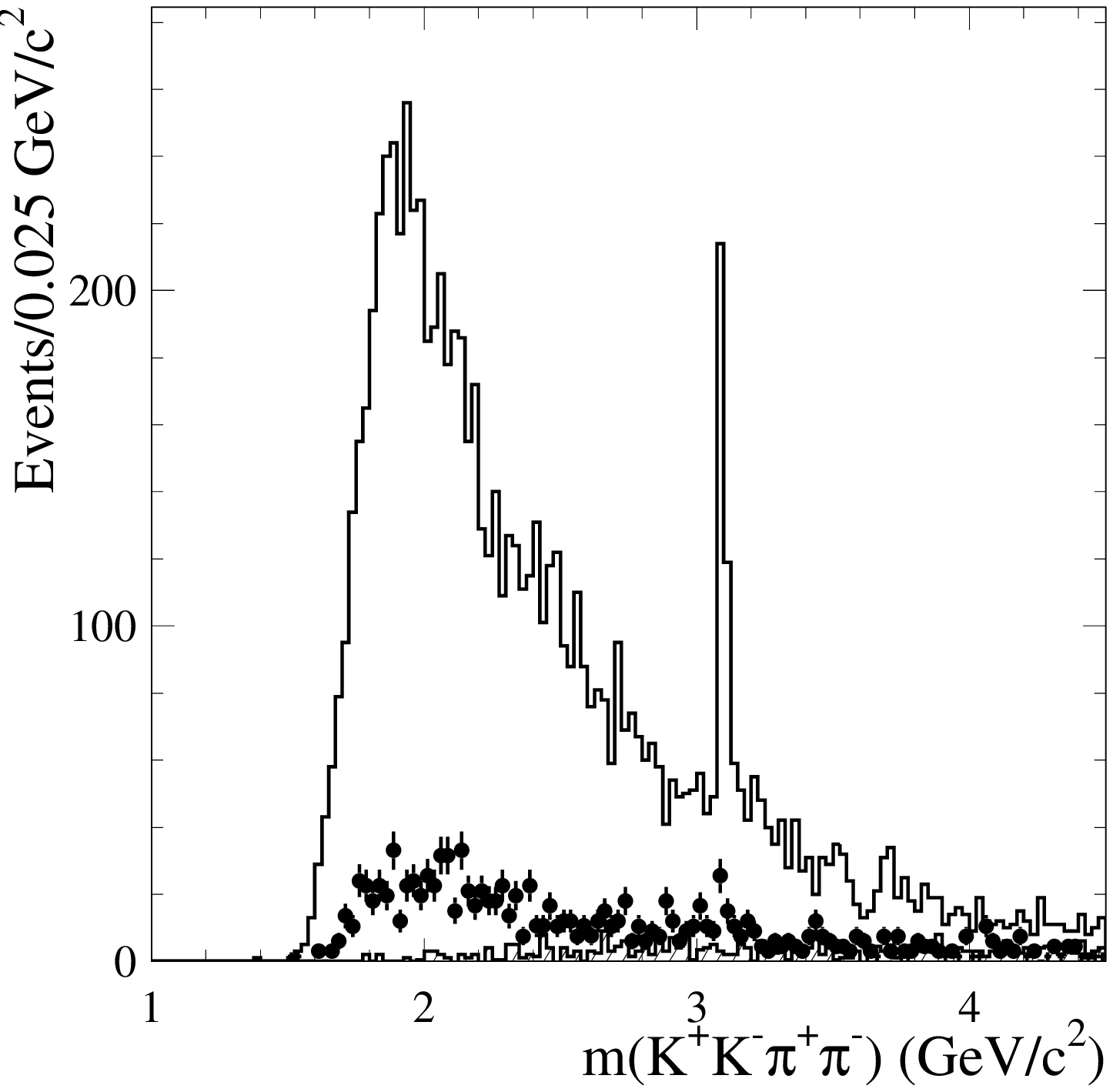}
\vspace{-0.3cm}
\caption{
The $K^+K^-\pipi$ invariant mass distribution for the signal
region of Fig.\@ 13. The points indicate the background estimated
       from the difference between data and MC simulation for the control region
       of Fig.\@ 13 normalized to the difference between the number of
data and MC events in the
       signal region of Fig.\@ 13. The shaded histogram corresponds
       to the non-ISR background estimated using JETSET.
}                          
\label{2k2pi}
\end{figure}
\begin{figure}[tbh]
\includegraphics[width=0.9\linewidth]{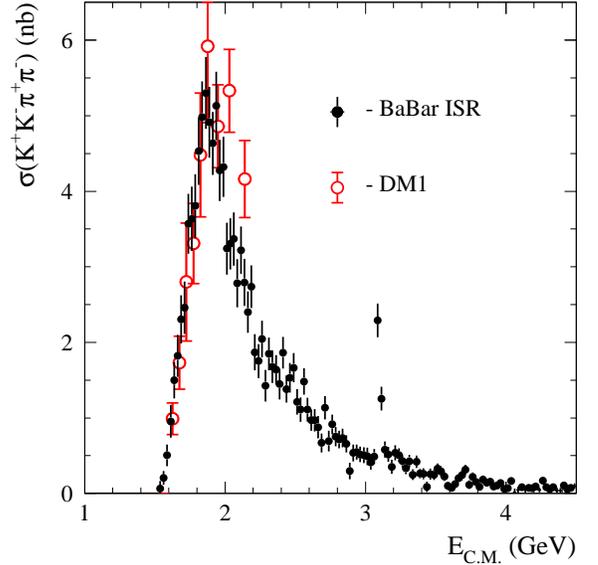}
\vspace{-0.3cm}
\caption{
The c.m.\@ energy dependence of the $\epem\to K^+K^-\pi^+\pi^-$
cross section obtained from ISR events at \babar\ compared with the only
direct \epem measurement, by DM1. Only statistical errors are shown.
}
\label{2k2pi_xs}
\end{figure} 

The selections $\chi^2_{2K2\pi}<20$, $\chi^2_{4\pi}>30$ and $\chi^2_{4K}>20$
are used, leaving a negligible number of $4K$ and 0.5\% of $4\pi$ events (from
simulation) in the final sample. The background subtraction procedure
which uses the control sample with
$20<\chi^2_{2K2\pi}<40$ is applied.  The background from
$4\pi$ events is not subtracted by this procedure, and 0.5\% of the events
shown in Fig.~\ref{4pi_babar} are used to make an additional correction.

In Fig.~\ref{2k2pi} we show the $K^+K^-\pipi$ invariant mass distribution
for the events that passed the selection procedure.  The points with error bars
show the background distribution from the control sample with $20<\chi^2_{2K2\pi}<40$
and the shaded histogram is the expected non-ISR background from 
JETSET MC simulation. Both are used for background subtraction. A very clear 
$J/\psi$ signal is seen.

Using the number of observed events, efficiency, and ISR luminosity, we
obtain the $\epem\to
K^+K^-\pi^+\pi^-$ cross section shown in
Fig.~\ref{2k2pi_xs}, which also displays the DM1 data~\cite{DM1}. 
Table~\ref{2k2pi_tab}
presents the cross section in 25~\mev bins, together with the ``undressed''
(no vacuum polarization) cross section.
The systematic errors are dominated by uncertainty in the acceptance 
simulation (10\%) and the difference between the kaon identification 
efficiencies for data and MC events (up to 
5\% per track), and are estimated to be 15\%. 

\begin{figure}
\includegraphics[width=0.95\linewidth,height=0.9\linewidth]{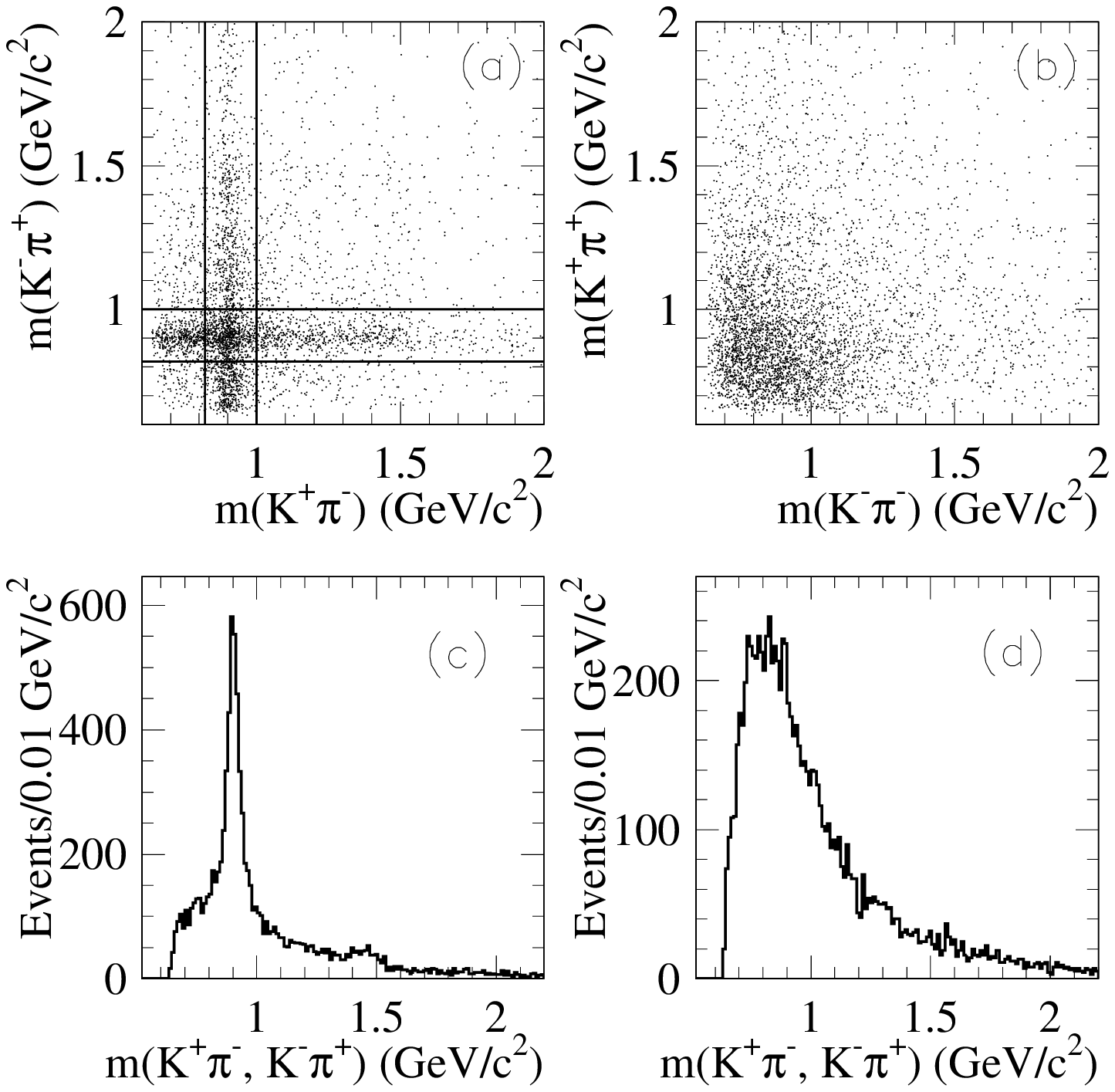}
\vspace{-0.3cm}
\caption{For the $K^+K^-\pi^+\pi^-$ data sample: (a) the scatter-plot of
the $K^+\pi^-$ and $K^-\pi^+$ invariant mass values; 
(b) the scatter-plot of the $K^+\pi^+$ and $K^-\pi^-$ invariant mass values;
(c) the $K^+\pi^-$ or $K^-\pi^+$ mass projection of (a);
(d) the mass distribution for other $K\pi$ combinations
for events in the $K^{*0}(892)$
bands of (a).
}
\label{2k2pi_mass1}
%
\vspace{-0.02cm}
\includegraphics[width=0.95\linewidth,height=0.9\linewidth]{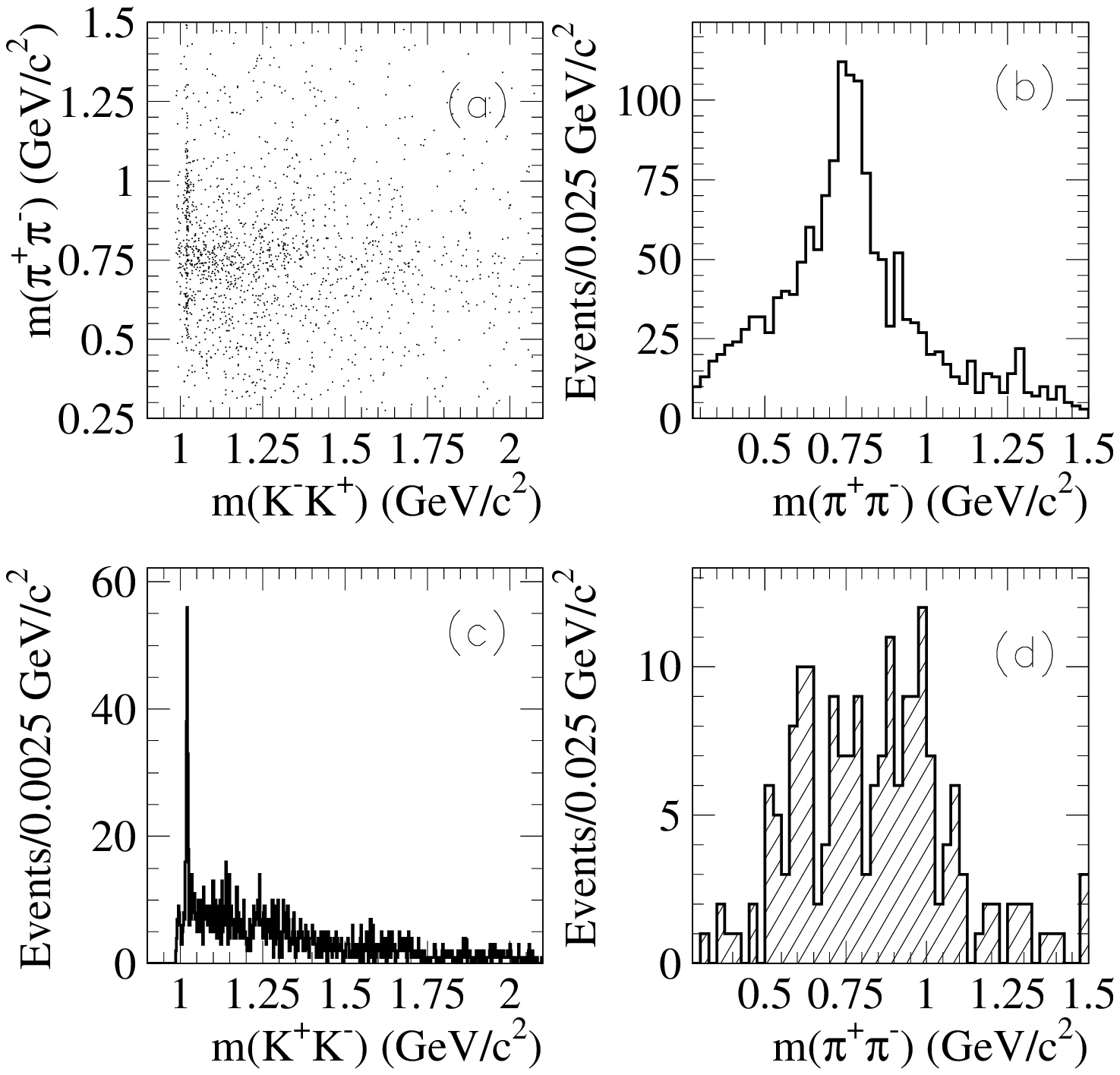}
\vspace{-0.3cm}
\caption{For the $K^+K^-\pi^+\pi^-$ data sample, after removing
events contributing to the $K^{*0}(892)$ mass regions indicated
in Fig.~\ref{2k2pi_mass1}(a):
(a) the scatter-plot of $\pi^+\pi^-$ and $K^+ K^-$ mass values;
(b) the $\pi^+\pi^-$ mass projection of (a);
(c) the $K^+ K^-$ mass projection of (a);
(d) the $\pi^+\pi^-$ mass projection of (a) for events 
in the $\phi(1020)$ mass peak.
}
\label{2k2pi_mass2}
\end{figure}

Figure~\ref{2k2pi_mass1} shows the $K\pi$ mass combinations.
Production of $K\pi$ pairs is dominated by the $K^{*0}(892)$
clearly seen in Fig.~\ref{2k2pi_mass1}(a).          
There also appears to be evidence of  $K^{*}_{2}(1430)$ production in
the $2K2\pi$ sample seen in the projection plot of
Fig.~\ref{2k2pi_mass1}(a) shown in Fig.~\ref{2k2pi_mass1}(c).
No structures in m($K^+\pi^+$) or in m($K^-\pi^-$)
are seen in Fig.~\ref{2k2pi_mass1}(b). 
The $K\pi$ mass distribution for the other $K\pi$ combination for
events in the $K^{*0}(892)$ bands of Fig.~\ref{2k2pi_mass1}(a) is shown
in Fig.~\ref{2k2pi_mass1}(d); events in the overlap region of the two
bands are included only once to avoid double-counting. The absence of
a clear $K^{*0}(892)$ or $K^{*}_{2}(1430)$ signal
 in Fig.~\ref{2k2pi_mass1}(d) indicates
that $K^{*0}(892)\Kbar^{*0}(892)$ and $K^{*0}(892) K^{*}_{2}(1430)$ 
quasi-two-body production reactions are small. 

When events in the $K^{*0}(892)$ bands of Fig.~\ref{2k2pi_mass1}(a)
are removed, the scatter-plot 
m($\pi^+\pi^-$) vs. m($K^+K^-$) in Fig.~\ref{2k2pi_mass2}(a)  shows
the presence of the $\rho^0$ and 
$\phi$ resonances. The $\pipi$
(Fig.~\ref{2k2pi_mass2}(b)) and $K^+ K^-$ (Fig.~\ref{2k2pi_mass2}(c)) mass
projections from Fig.~\ref{2k2pi_mass2}(a) exhibit clear
$\rho^0$ and $\phi$ signals, respectively.
Figure~\ref{2k2pi_mass2}(d) shows the $\pi^+\pi^-$
mass distribution for events with m($K^+K^-$) in the $\phi$ region. 
The absence of any $\rho$ signal is consistent with $C$ conservation;
there might be slight evidence of $f_{0}(980)$ production.

\begin{figure}
\includegraphics[width=0.95\linewidth,height=0.9\linewidth]{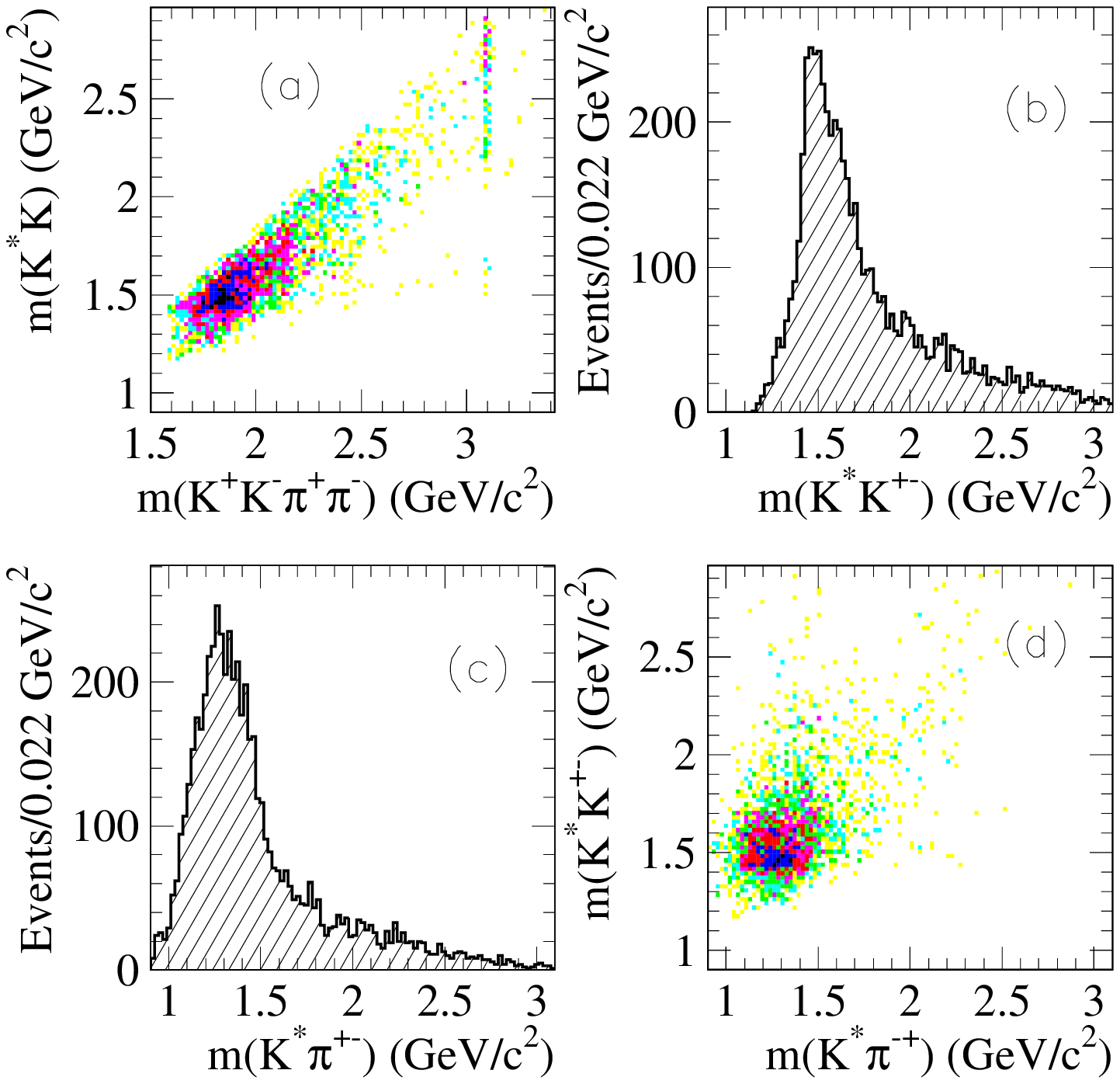}
\vspace{-0.3cm}
\caption{The $2K2\pi$ events from the $K^{*0}(892)$ bands in 
Fig.~\ref{2k2pi_mass1}(a): 
(a) the $K^{*0}(892)K^\pm$ vs. $K^+K^-\pi^+\pi^-$ invariant mass
distribution scatter-plot; (b) the $K^{*0}(892)K^\pm$ mass projection; 
(c) the $K^{*0}(892)\pi^{\pm}$ mass distribution; (d) the $m(K^{*0}\pi^{\mp})$
vs. $m(K^{*0}K^\pm)$ scatter-plot.
}
\label{2k2pi_mass3}
\vspace{-0.02cm}
\includegraphics[width=0.95\linewidth,height=0.9\linewidth]{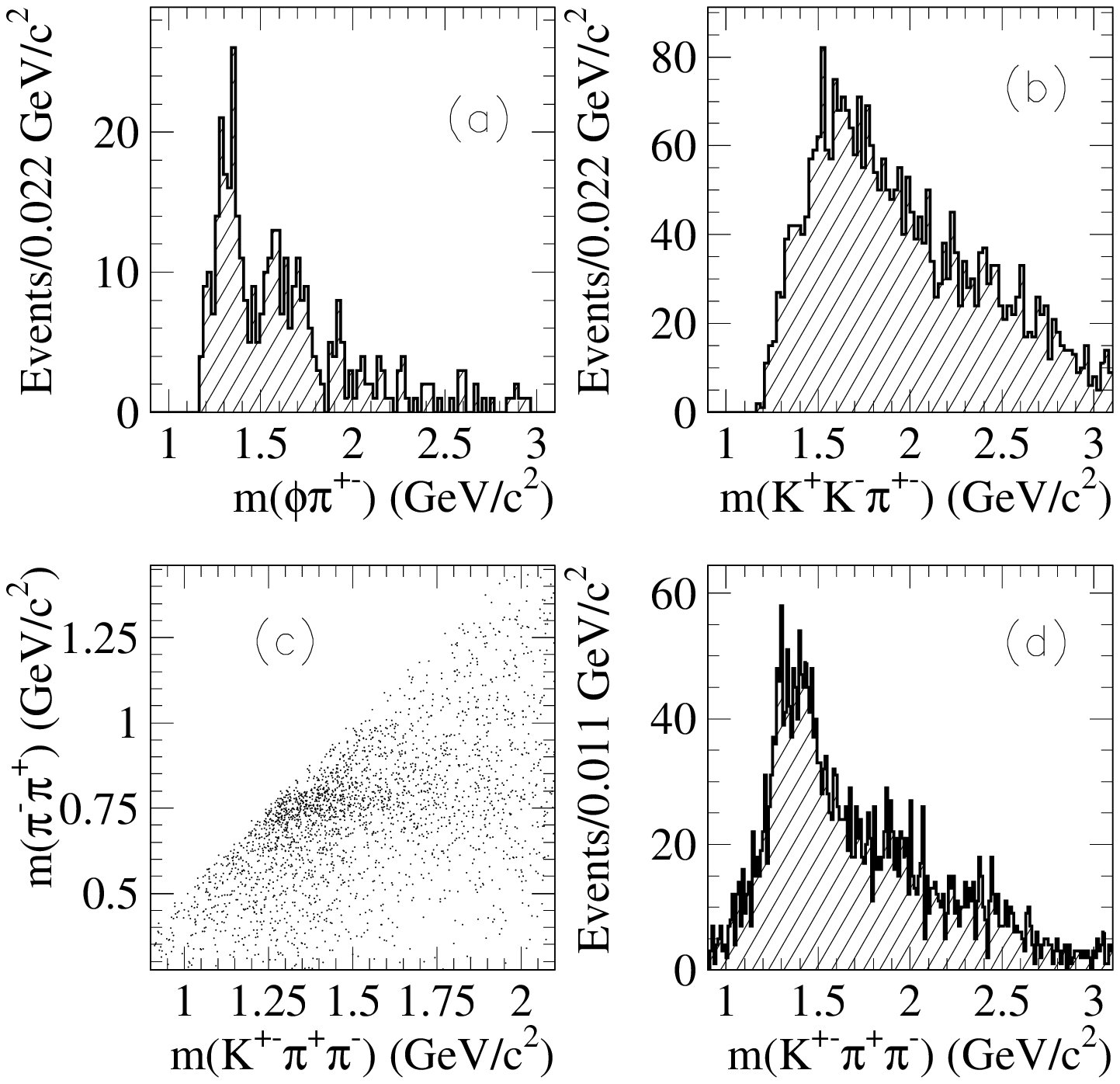}
\vspace{-0.3cm}
\caption{The $2K2\pi$ events with the $K^{*0}(892)$ bands in
  Fig.~\ref{2k2pi_mass1}(a) excluded:
(a) the $K^+K^-\pi^\pm$ mass distribution with the $K^+K^-$ mass in
the  $\phi$ region;
(b) the $K^+K^-\pi^\pm$ mass distribution with 
$K^+K^-$ in the $\phi$ mass region excluded;
(c) the $\pi^+\pi^-$ vs. $K^\pm\pi^+\pi^-$ scatter-plot, and 
(d) the $K^\pm\pi^+\pi^-$ projection, for the events from (b).
}
\label{2k2pi_mass4}
\end{figure}

The three-body mass combinations are also of potential interest. We consider
these in two categories. For the first category, we require that there
be a  $K^{\mp}\pi^\pm$
combination in one of the $K^{*0}(892)$ bands of Fig.~\ref{2k2pi_mass1}(a).
The $K^{*0}(892)K^\pm$ invariant mass behavior for such events is shown in
Figs.~\ref{2k2pi_mass3}(a,b). In Fig.~\ref{2k2pi_mass3}(b), there is a strong
peak just below 1.5~\gevcc followed by a shoulder around 1.6--1.7~\gevcc and
a rapid drop in the 1.8--2.0~\gevcc region. The peak could correspond to a
presently unknown isovector state decaying to $K^{*0}\Kbar + \Kbar^{*0}K$.
The corresponding $K^{*0}\pi$ behavior is shown in
Figs.~\ref{2k2pi_mass3}(c,d). The mass projection in Fig.~\ref{2k2pi_mass3}(c)
shows a broad structure in the region of the $K_{1}(1270)$ and
$K_{1}(1410)$, both of which are known to decay through  $K^{*0}\pi$.
The scatter-plot of m$(K^{*0}\pi)$ vs. m$(K^{*0} K)$ in Fig.~\ref{2k2pi_mass3}(d)
shows that the low-mass enhancements of
Figs.~\ref{2k2pi_mass3}(b,c) are
highly-correlated. 

For the second category, we exclude events with at least one $K \pi$
combination in  a  $K^{*0}(892)$ band of Fig.~\ref{2k2pi_mass1}(a). As shown in
Fig.~\ref{2k2pi_mass2}(c), a subset of the remaining events contains a
$\phi$ signal. When the events in the $\phi$ region are combined with
the remaining $\pi^\pm$, the $\phi\pi^\pm$ mass distribution of
Fig.~\ref{2k2pi_mass4}(a) is obtained. This shows a peak in the
mass region 1.25--1.4~\gevcc, followed by a second broad
bump in the 1.5--1.8~\gevcc region. If the $\phi$ region is
excluded, the rather featureless mass distribution of
Fig.~\ref{2k2pi_mass4}(b) is obtained. However, the scatter-plot of
m$(\pipi)$ vs. m$(K^\pm\pipi)$ in Fig.~\ref{2k2pi_mass4}(c) for these events
shows clear evidence for $\rho$ production
correlated with $K^\pm\pipi$ mass in the 1.2--1.5~\gevcc
region. The $K_{1}(1270)$ couples quite strongly to $K\rho$ ~\cite{PDG}, and so
may be produced in this final state. However, the $K_{1}(1410)$ is
almost decoupled from $K\rho$~\cite{PDG}, and so cannot be the source of the
events in the 1.4--1.5~\gevcc region.

There is evidence in the $K^+K^-\pi^+\pi^-$ final state for
several interesting, but complex, structures. Detailed understanding
will require data on $K^+K^-\pi^0\pi^0$ and also on four-body final
states involving neutral kaons. 
%
%
\section{\boldmath The $K^+K^-K^+K^-$ final state}
\begin{figure}
\includegraphics[width=0.9\linewidth]{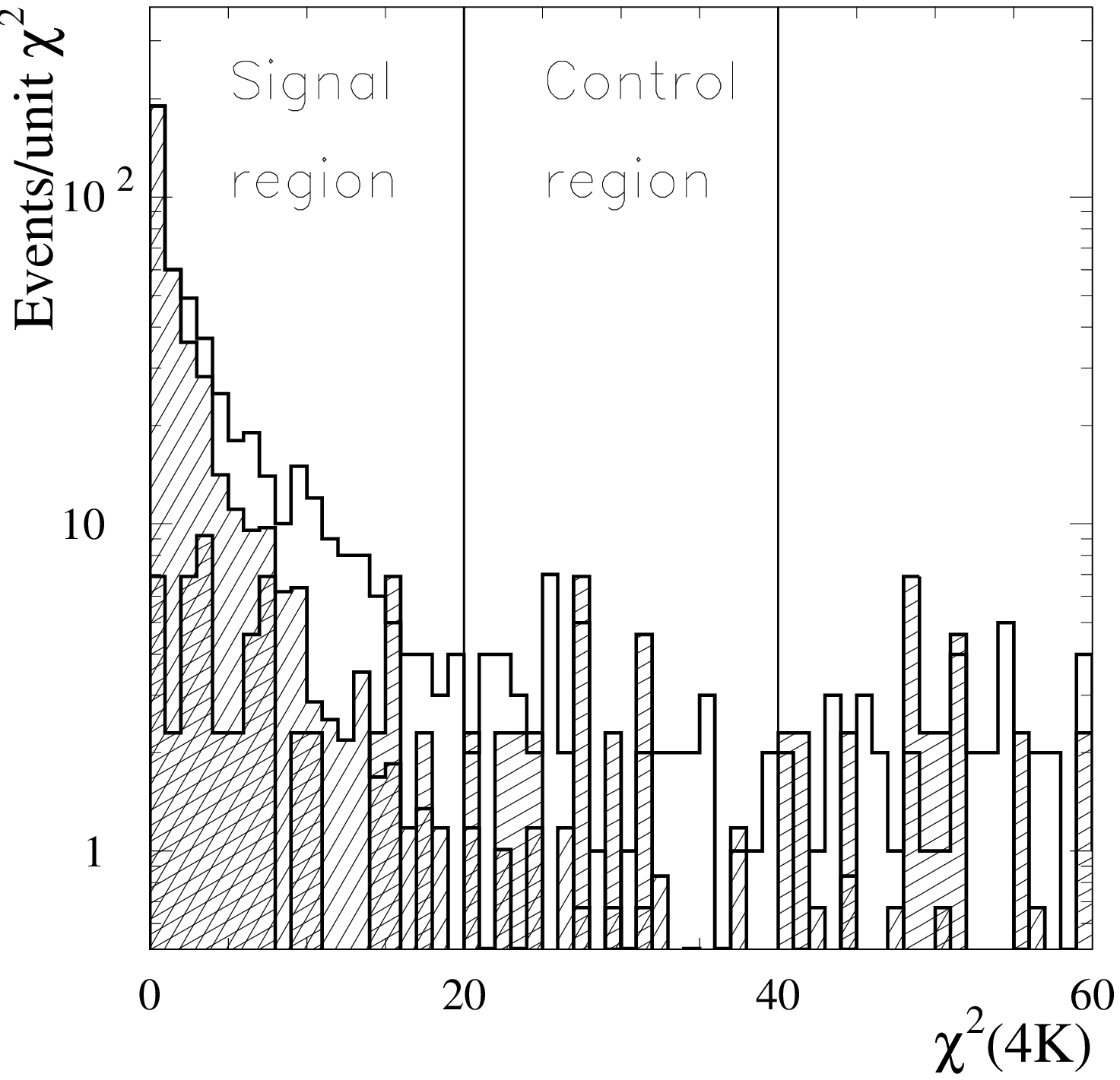}
\vspace{-0.3cm}
\caption{
      The one-constraint \chisq  distributions for
       four-charged-track data events and $K^+ K^- K^+ K^-$ Monte Carlo
       events (shaded histogram) fitted to the $K^+ K^- K^+ K^-$ 
       hypothesis. At least three kaons must be identified. The
      cross-hatched  histogram is the estimated background contribution from 
       $K^+ K^-\pi^+\pi^-$ ISR events. The signal and control regions 
       are indicated.
}
\label{chi2_4k}
\includegraphics[width=0.9\linewidth]{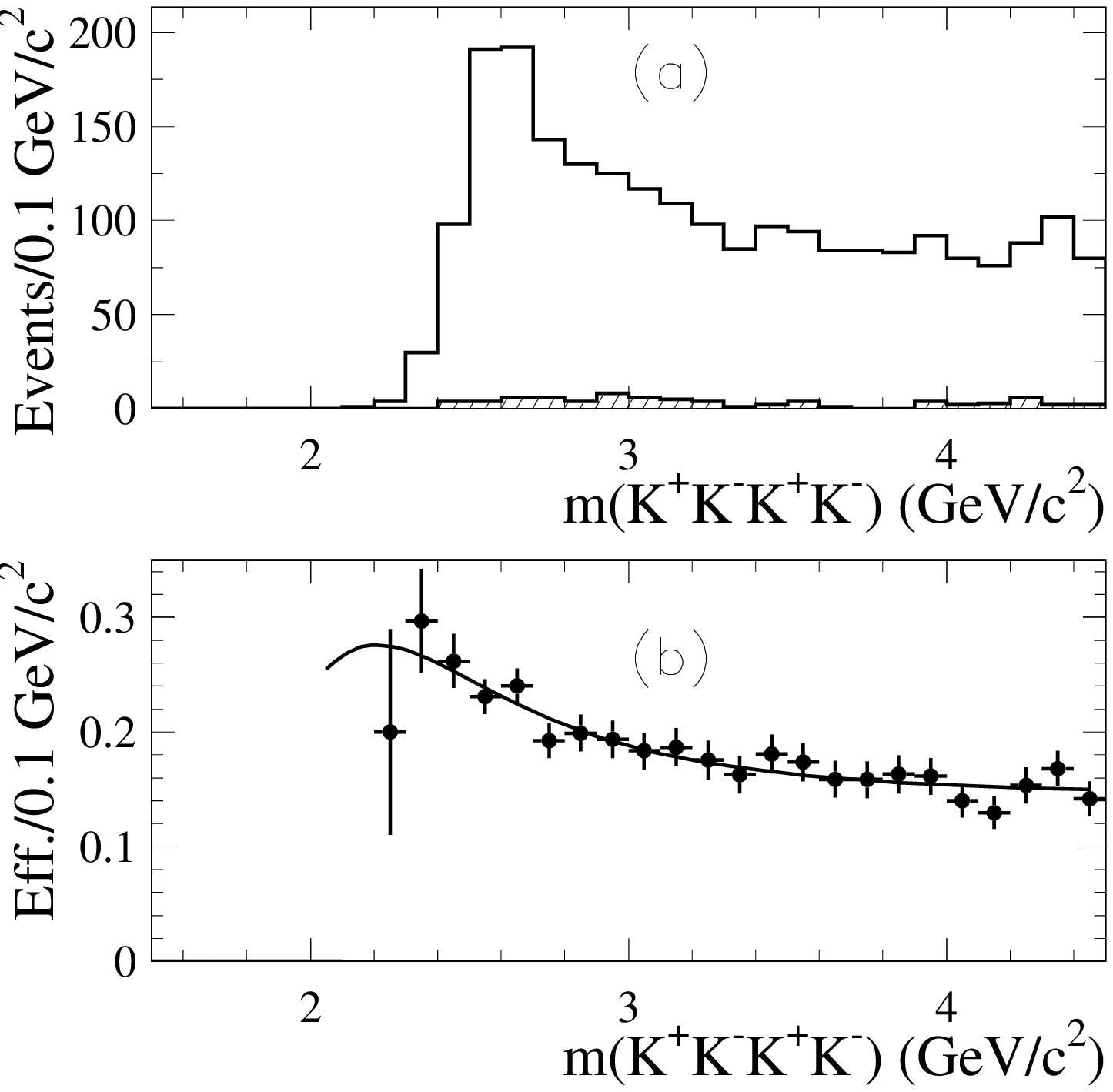}
\vspace{-0.5cm}
\caption{
       (a) The $K^+ K^- K^+ K^- $ mass distribution from simulation for 
           the signal and control (shaded) regions of Fig.~\ref{chi2_4k}; (b) the 
           mass dependence of the net reconstruction and selection 
           efficiency obtained from simulation.
}
\label{4k_acc}
\end{figure} 
\begin{figure}
\includegraphics[width=0.9\linewidth]{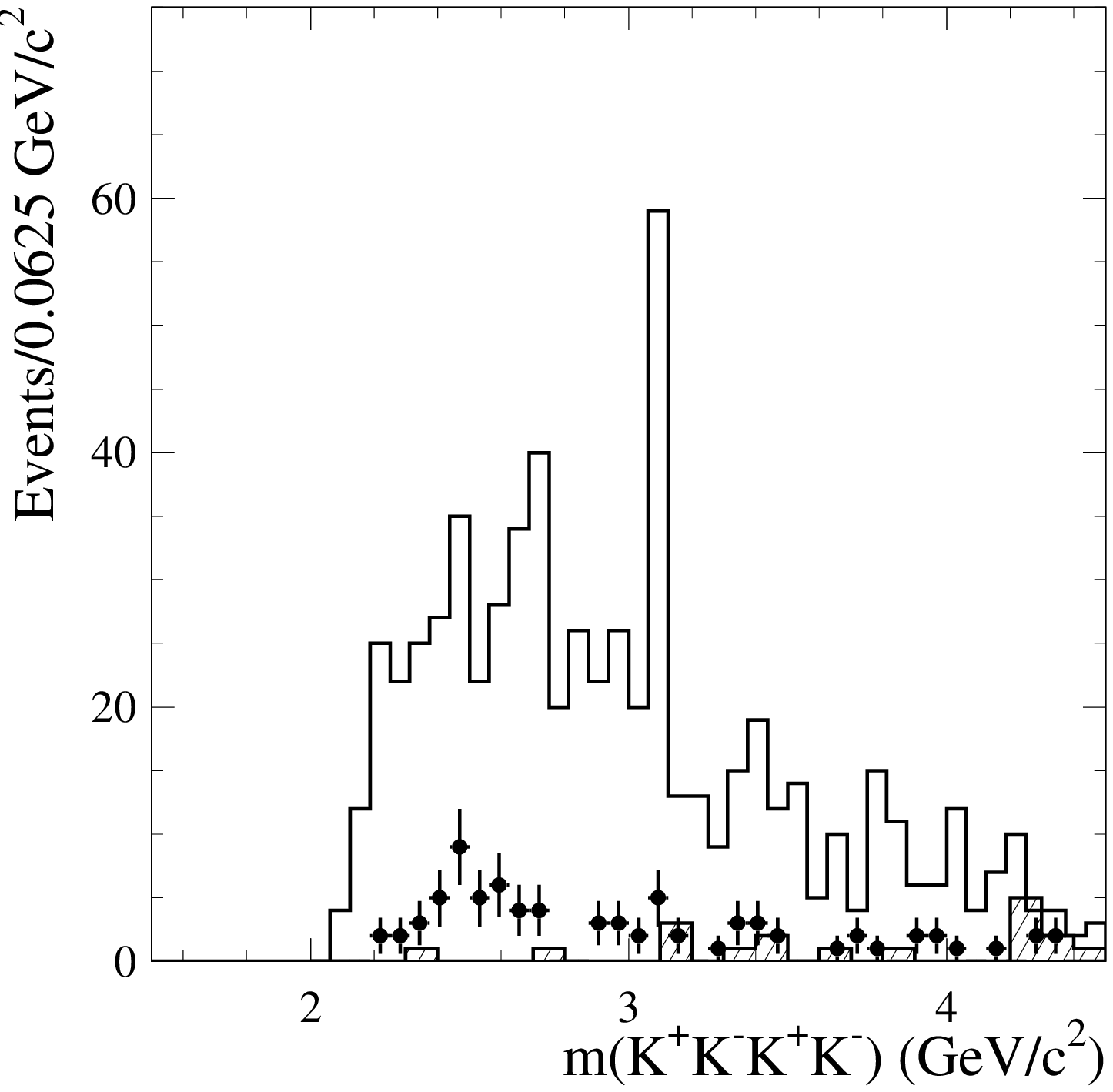}
\vspace{-0.3cm}
\caption{
The $K^+K^-K^+K^-$ invariant-mass distribution for the signal
region of Fig.~\ref{chi2_4k}. The points indicate the background estimated
from the difference between data and MC simulation for the control region
of Fig.~\ref{chi2_4k}, normalized to the difference between the number
of data and MC events in the
signal region of Fig.~\ref{chi2_4k}. The shaded histogram corresponds
to the non-ISR background estimated from JETSET.
}
\label{4k_all}
%
\includegraphics[width=0.9\linewidth]{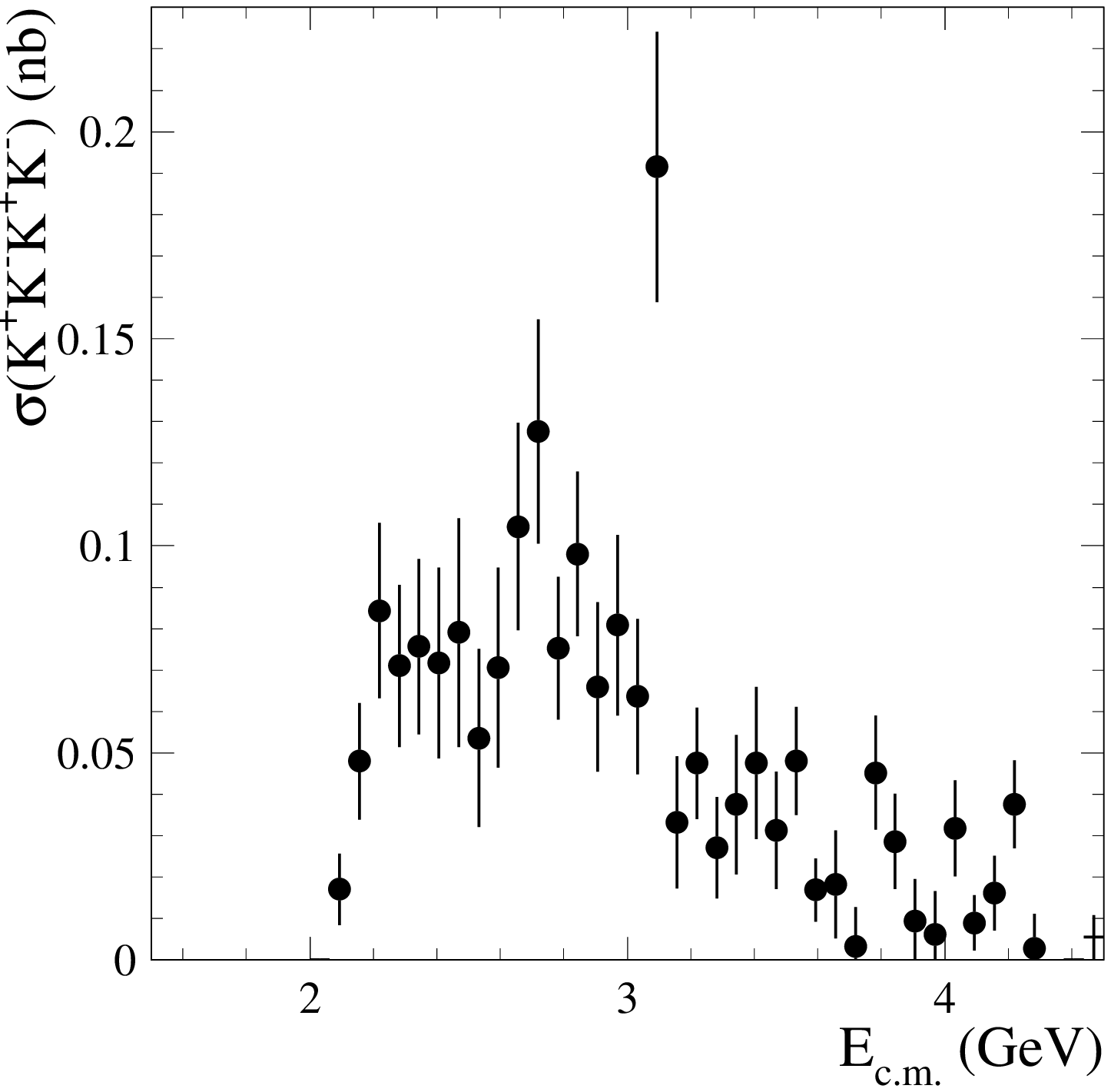}
\vspace{-0.3cm}
\caption{
The c.m.\@ energy dependence of the $\epem\to K^+K^-K^+K^-$
cross section obtained from ISR events at \babar. Only statistical errors are 
shown.
}
\label{4k_xs}
\end{figure} 
\begin{figure}[tbh]
\includegraphics[width=\linewidth]{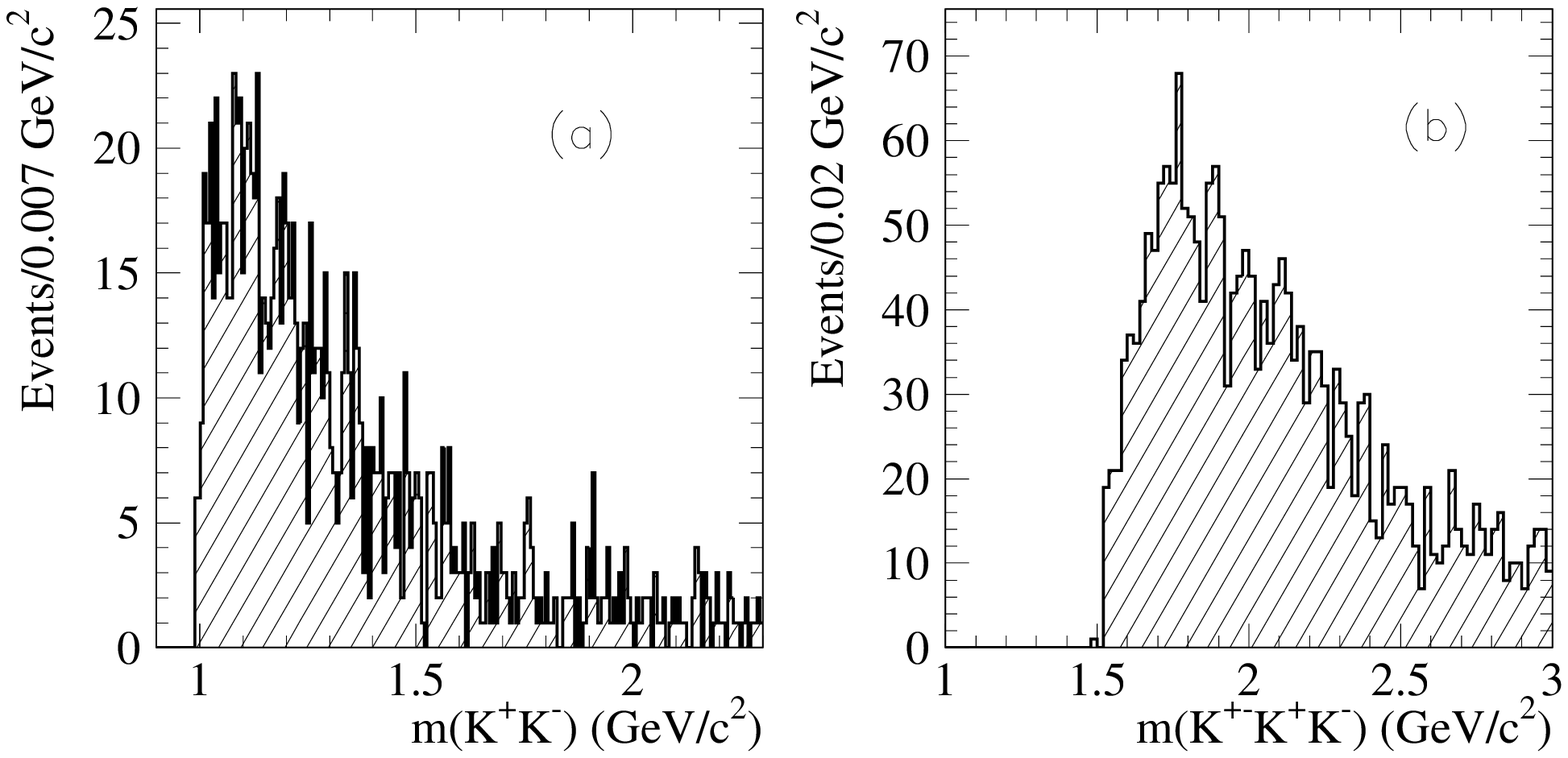}
\vspace{-0.3cm}
\caption{
(a) The $K^+K^-$ and (b) the $K^\pm K^+K^-$  invariant mass 
distributions for events from the $K^+K^-K^+K^-$ sample.
}
\label{2k3k_mass}
\end{figure}  
The one-constraint fit for the four-charged-kaon hypothesis 
with $\chi^2_{4K}<20$ selection gives a very pure sample
of this final state. The main background contribution is from the
reaction with two charged kaons and two misidentified charged pions.
We reduced this background greatly by adding to the initial selection criterion
the requirements that three or four of
the charged particles have good kaon identification and that 
$\chi^2_{2K2\pi}>20$ for the fit to the reaction with two kaons and two pions.
Background from the $4\pi$ final state is negligible.

Figure~\ref{chi2_4k} shows the \chisq distributions for data and simulation.
The simulation of the $4K$ final state uses a point-like matrix element
with cross section behavior close to what we observe experimentally,
and with all radiative
processes included.  Also shown is the contribution from the $2K2\pi$ final
state obtained from the simulation (cross-hatched).  

Figure~\ref{4k_acc}(a) shows the simulated mass distribution for $4K$
events, and
Fig.~\ref{4k_acc}(b) presents the detection efficiency calculated as
the ratio of selected to generated $4K$ MC events.

In Fig.~\ref{4k_all} the $K^+K^-K^+K^-$ invariant mass distribution is shown
for the data in the signal region, and also for background estimated from the
$20<\chi^2_{4K}<40$ control sample (points). The shaded histogram shows the
expected non-ISR background from the JETSET MC simulation. A clear signal due to the $J/\psi$
is seen. The number of signal events as a function of four-kaon mass is
obtained by subtracting the background contributions estimated from the
control sample and from the JETSET MC simulation.

Using the number of four-charged-kaon events observed, the acceptance
discussed previously, and the ISR luminosity, we calculate the cross section for the
reaction $\epem\to K^+K^-K^+K^-$. The results with statistical errors
only are displayed in
Fig.~\ref{4k_xs}, and the values are listed in Table~\ref{4k_tab}.  There are
no published electron-positron data for comparison.  Systematic errors are
dominant for these measurements. These uncertainties are due to the
absence  of a detailed
model for the acceptance simulation, to uncertainties in the
background  subtractions, and
to differences in kaon identification efficiency in the MC simulation
and the data. The overall
estimated systematic error is about 25\%. The $K^+K^-$ and $K^\pm K^+K^-$
invariant mass distributions are shown in Fig.~\ref{2k3k_mass} (four entries
per event). If there is any $\phi$ production it is small, and
neither mass distribution shows  evidence of significant structure.
\section{\boldmath The $J/\psi$ region}
\begin{figure}[tbh]
\includegraphics[width=0.95\linewidth]{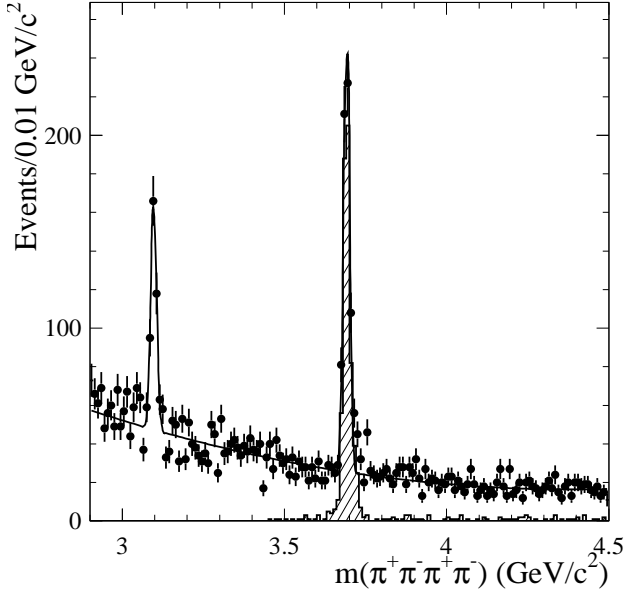}
\caption{
The $\pipi\pipi$ mass distribution for ISR-produced
$\epem\to\pipi\pipi$ events in the $J/\psi$--$\psi(2S)$
        region; there are clear signals at the $J/\psi$ and 
        $\psi(2S)$ mass positions. The shaded region at the 
        latter corresponds to $\psi(2S)\to J/\psi\pipi$, with 
        $J/\psi\to\mu^+\mu^-$ and the muons are treated as pions.
}
\label{jpsi}
\end{figure}

Figure~\ref{jpsi} shows an expanded view of the $J/\psi$ mass region 
in Fig.~\ref{4pi_babar} for the four-pion data sample. 
The signals from $J/\psi\to\pipi\pipi$ and $\psi(2S)\to 
J/\psi\pipi\to\mumu\pipi$ (without muon identification) are clearly seen. 
\begin{figure}[tbh]
\includegraphics[width=0.9\linewidth,height=0.85\linewidth]{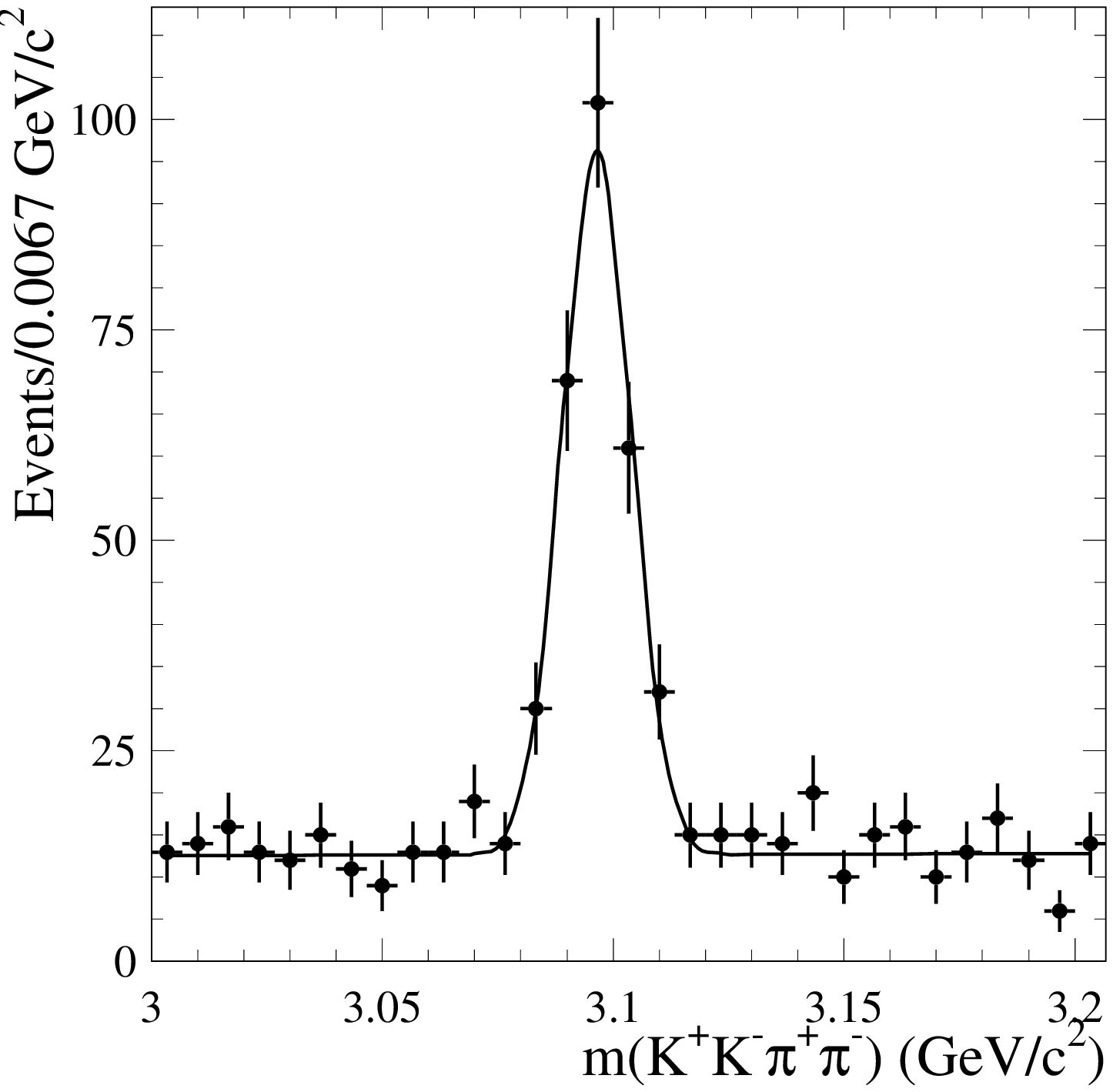}
\includegraphics[width=0.9\linewidth,height=0.85\linewidth]{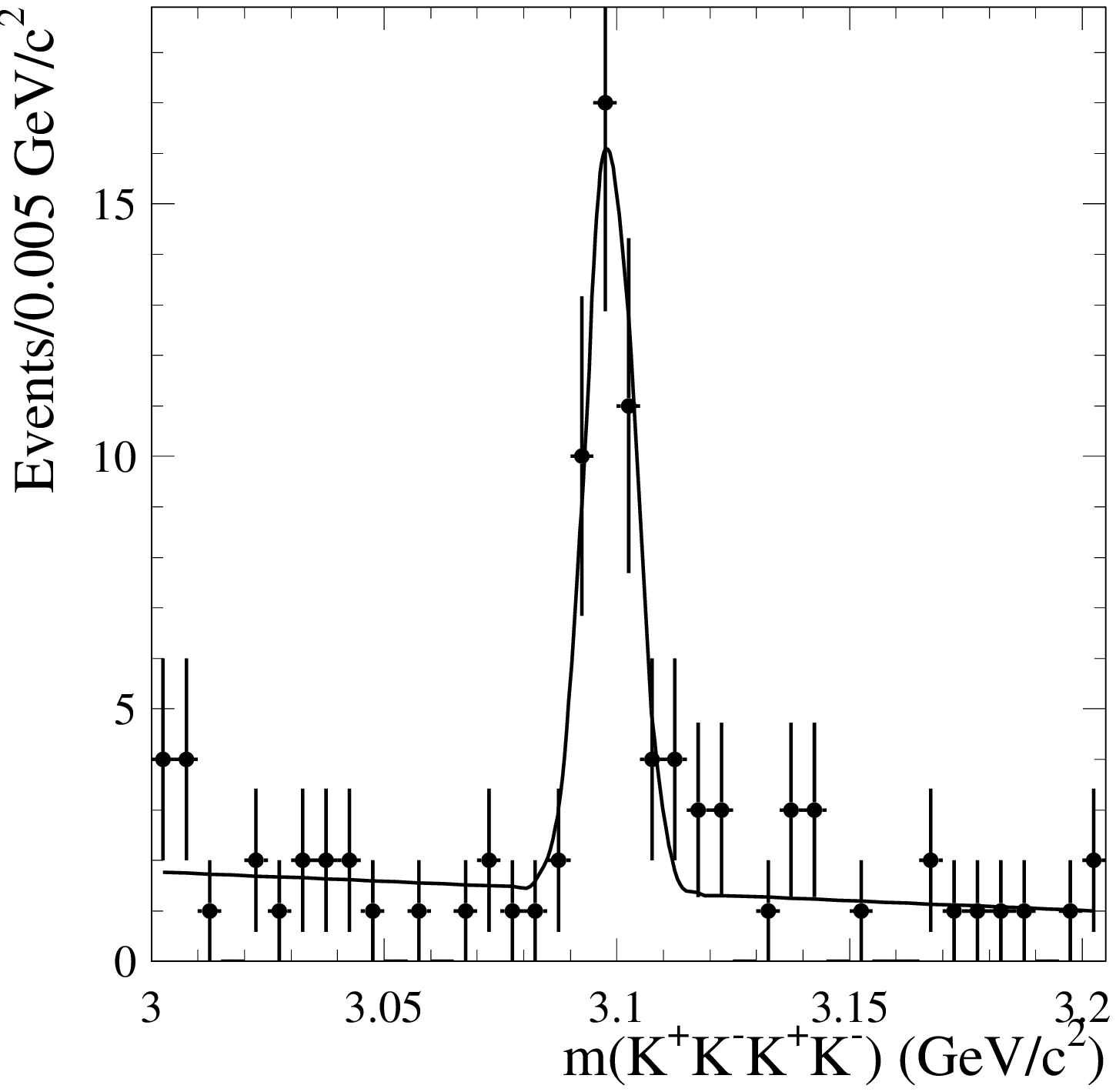}
\caption{
The $K^+K^-\pi^+\pi^-$ (top) and $K^+ K^- K^+ K^-$ (bottom) mass 
distributions for ISR-produced events in the $J/\psi$ region.
        A clear signal is observed at the $J/\psi$ mass.
}
\label{jpsi_2k2pi}
\end{figure}
The observation of $J/\psi$ decaying into four pions gives a direct
measurement of the 4$\pi$-mass resolution and the absolute energy scale. A
fit with a Gaussian for the $J/\psi$ peak and a polynomial function
for the continuum, gives
$\sigma_{m(4\pi)}=8~\mevcc$ and less than 1~\mevcc difference
from the PDG~\cite{PDG} value for the $J/\psi$ mass. The observed mass resolution
agrees with the simulation within 10\%.

The observed $270\pm20$ events at the $J/\psi$ peak can be used to calculate
the branching fraction for $J/\psi\to\pipi\pipi$.  The simulation shows that
because of radiative effects only 89\% of the signal events are under the
Gaussian curve. This value is in good agreement with that obtained from
$\psi(2S)$ peak 87.7$\pm$1.3\% (see below) which is used for calculations.
Using the corrected number, we can calculate the products:
\begin{eqnarray*}
  B_{J/\psi\to 4\pi}\cdot\sigma_{\rm int}^{J/\psi}
  &=& \frac{N(J/\psi\to\pi^+\pi^-\pi^+\pi^-)}%
           {d{\cal L}/dE\cdot\epsilon_{\rm MC}} \\
  &=& 46.8\pm3.5\pm3.3~\nb\mev\ ,\\
  B_{J/\psi\to 4\pi}\cdot\Gamma^{J/\psi}_{ee}
  &=& \frac{N(J/\psi\to\pipi\pipi)\cdot m_{J/\psi}^2}%
           {6\pi^2\cdot d{\cal L}/dE\cdot\epsilon_{\rm MC}\cdot C} \\
  &=& (1.95\pm0.14\pm0.13)\times 10^{-2}~\kev\ ,\\
\end{eqnarray*}
where
$$\sigma_{\rm int}^{J/\psi} = 6\pi^2\Gamma^{J/\psi}_{ee}C/m_{J/\psi}^2 =
  12790\pm850 \mbox \nb\cdot\mev$$
is the integral over the $J/\psi$ excitation curve; $\Gamma^{J/\psi}_{ee}$ is
the electronic width; $d{\cal L}/dE = 25.3~\invnb/\mev$ is the ISR luminosity
at the $J/\psi$ mass; $\epsilon_{\rm MC} = 0.26\pm0.01$ is the detection
efficiency from simulation with the corrections discussed in 
Sec.~\ref{sec:Systematics};
and  $C = 3.894\times 10^{11}~\nb\mev^2$ is a
conversion constant.  The systematic error includes 3\% uncertainty from
the ISR luminosity, 3\% from efficiency, and 5\% uncertainty from background
subtraction.  The subscript ``$4\pi$'' for branching fractions refers
to the $\pipi\pipi$ final state exclusively.

Using $\Gamma^{J/\psi}_{ee} =5.40\pm0.17~\kev$ ~\cite{PDG}, we obtain the
result $B_{J/\psi\to 4\pi} = (3.61\pm 0.26\pm 0.26)\times 10^{-3}$, 
substantially more precise that the current PDG value
$B_{J/\psi\to 4\pi} = (4.0\pm 1.0)\times 10^{-3}$ ~\cite{PDG}. The systematic error
includes 3\% uncertainty in $\Gamma^{J/\psi}_{ee}$.

The $\psi(2S)$ peak corresponds to the decay chain
$\psi(2S)\to J/\psi\pipi\to\mumu\pipi$ with the muons treated as
pions.  The number of events extracted from a fit to a Gaussian
distribution for the $\psi(2S)$ peak and
a polynomial function for the continuum is $544\pm 27$. With a radiative
correction of 0.89 taken from simulation, $N(\psi(2S)\to J/\psi\pipi) =
611\pm30$ events is the observed signal.

The number of $\psi(2S)$ events can be obtained with much
less background if the invariant mass of one pair of charged tracks 
assumed to be muons is
within $\pm 50~\mevcc$ of the $J/\psi$ mass.
Events satisfying this criterion
are shown by the shaded histogram 
in Fig.~\ref{jpsi}; this $\psi(2S)$ peak has $620\pm 25$ events after
subtraction of 10 background events. 
With this number and the $544\pm 27$ events obtained from the Gaussian fit, the
radiative correction value can be obtained directly; the value 
is $0.877\pm 0.013$, in good agreement with the value 0.89 from 
simulation.  

For the $\psi(2S)$ we then obtain:
\begin{eqnarray*}
  \lefteqn{B_{\psi(2S)\to J/\psi\pipi} \cdot B_{J/\psi\to \mumu}
      \cdot\sigma_{\rm int}^{\psi(2S)}} \\
  &\qquad =& \frac{N(\psi(2S)\to J/\psi\pi^+\pi^-)}
      {d{\cal L}/dE\cdot\epsilon_{\rm MC}} \\
  &\qquad =& 76.3\pm3.1\pm3.8~\nb\mev
\end{eqnarray*}
\noindent which leads to\\
\begin{eqnarray*}
  \lefteqn{B_{\psi(2S)\to J/\psi\pi^+\pi^-} \cdot B_{J/\psi\to\mumu}
    \cdot\Gamma^{\psi(2S)}_{ee}} \\
  &\qquad =& \frac{N(\psi(2S)\to J/\psi\pi^+\pi^-)\cdot m_{\psi(2S)}^2}
           {6\cdot\pi^2\cdot d{\cal L}/dE\cdot\epsilon_{\rm MC}\cdot C} \\
  &\qquad =& (4.50\pm0.18\pm0.22)\times 10^{-2}~\kev\ ,
\end{eqnarray*}
where $d{\cal L}/dE = 32.4~\invnb/\mev$ is the ISR luminosity at
the $\psi(2S)$ mass, and $\epsilon_{\rm MC} = 0.25\pm 0.01$ is the detection 
efficiency from simulation with corrections discussed in 
Sec.~\ref{sec:Systematics}. 
The same systematic errors as for the
$J/\psi$ have been included. 

Using the values $\Gamma^{\psi(2S)}_{ee} = 2.12\pm 0.12$~\kev and
$B_{J/\psi\to \mumu} = 0.0588\pm 0.0010$ from Ref.~\cite{PDG}, we obtain
$$B_{\psi(2S)\to J/\psi\pi^+\pi^-} = 0.361\pm 0.015\pm 0.028\ ,$$
which should be compared to the current world-average value, $B_{\psi(2S)\to
J/\psi\pi^+\pi^-} = 0.317\pm 0.011$~\cite{PDG}. Almost half of our
systematic error in $B_{\psi(2S)\to J/\psi\pi^+\pi^-}$ comes from
the uncertainty in $\Gamma^{\psi(2S)}_{ee}$.

Alternatively, using the PDG values for the two branching fractions, we can
extract the \epem width of the $\psi(2S)$,
$$\Gamma^{\psi(2S)}_{ee} = 2.41\pm 0.10\pm 0.15~\kev\ ,$$
where the systematic error has been increased due to the uncertainty
in $B_{\psi(2S)\to J/\psi\pi^+\pi^-}$.

Figure~\ref{jpsi_2k2pi} shows the $J/\psi$ signal in the
$K^+ K^-\pipi$ and $K^+ K^- K^+ K^- $ modes.  The numbers of events under the 
Gaussian curves
are $233\pm 19$ and $38.5\pm 6.7$ respectively. The mass resolution is
about 7~\mevcc for $2K2\pi$ and 5~\mevcc for the $4K$
final state.  Using the radiative correction factor 0.877, and
$\epsilon_{2K2\pi} = 0.13\pm 0.01$ and $\epsilon_{4K} = 0.18\pm 0.02$ from
simulation, we obtain
\begin{eqnarray*}
  B_{J/\psi\to 2K2\pi}\cdot\Gamma^{J/\psi}_{ee}
  &=& (3.36\pm0.27\pm0.27)\times 10^{-2}~\kev \\
  B_{J/\psi\to 4K}\cdot\Gamma^{J/\psi}_{ee}
  &=& (4.0\pm0.7\pm0.6)\times 10^{-3}~\kev\ .
\end{eqnarray*}
The systematic errors are mainly due to the uncertainties in acceptance and
ISR luminosity. 

Using the PDG value for $\Gamma^{J/\psi}_{ee}$, we calculate the branching fractions\\
\begin{eqnarray*}
  B_{J/\psi\to 2K2\pi} &=& (6.2\pm 0.5\pm 0.5)\times 10^{-3}\\
  B_{J/\psi\to 4K} &=& (7.4\pm 1.2\pm 1.2)\times 10^{-4}\ ,\\ 
\end{eqnarray*}
to be compared with the current PDG values of ($7.2\pm 2.3$)$\cdot$10$^{-3}$
and ($9.2\pm 3.3$)$\cdot$10$^{-4}$ respectively. The uncertainty in
$\Gamma^{J/\psi}_{ee}$ has been added in quadrature to the systematic error estimate.

Since we have measured the products of $\sigma_{\rm int}^{J/\psi}$ and 
branching fraction
of $J/\psi$ decay to $4\pi$, $2K2\pi$, and $4K$ 
it is interesting to compare them with the non-resonant cross sections 
(continuum) at that energy.
Using a linear approximation of the cross sections from
Tables~\ref{4pi_tab}, \ref{2k2pi_tab}, \ref{4k_tab} around the $J/\psi$
peak within $\pm 0.1\gev$
(events from the peak are excluded), the following cross
sections are obtained at the $J/\psi$ mass:
\begin{eqnarray*}
  \sigma_{4\pi}   &=& 0.55\pm0.03~\nb \\
  \sigma_{2K2\pi} &=& 0.48\pm0.04 \nb \\
  \sigma_{4K}     &=& 0.055\pm0.009 \nb\ .
\end{eqnarray*}
Table~\ref{jpsi_to_xs} presents the
ratios $B_{J/\psi\to f}\cdot\sigma_{\rm int}^{J/\psi}/\sigma_{\epem\to f}$ for $f=4\pi,
2K2\pi, 4K$. In these ratios all experimental systematic errors cancel. 
Also shown is the ratio 
$B_{J/\psi\to\mumu}\cdot\sigma_{\rm int}^{J/\psi}/\sigma_{\epem\to \mumu}$ 
taken from Ref.~\cite{Druzhinin1}.
\begin{table}[tbh]
\caption{
Ratios of the $J/\psi$ partial production rates to continuum cross
sections. The result for $\mumu$ is from Ref.~\cite{Druzhinin1}.
}
\label{jpsi_to_xs}
\begin{ruledtabular}
\begin{tabular}{cc}
Final state, $f$ & 
$B_{J/\psi\to f}\cdot\sigma_{\rm int}^{J/\psi} / \sigma_{\epem\to f}$
(MeV) \\ \hline
$\pipi\pipi$ & $85.1 \pm 7.9$ \\
$K^+ K^-\pipi$ & $166 \pm 19$ \\ 
$K^+ K^- K^+ K^-$ & $138 \pm 32$ \\
$\mumu$ & $84.12 \pm 0.67$ 
\end{tabular}
\end{ruledtabular}
\end{table}

The ratio obtained for the $4\pi$ final state is in good agreement
with that for $\mumu$.
Indeed, the strong decay of the $J/\psi$ to $4\pi$ is forbidden by G-parity conservation
and therefore this decay is expected to be dominated by single-photon.
No such suppression due to G-parity for the strong decay of the $J/\psi$
for the other two modes.
We interpret the significantly larger values of the ratios as an indication that
the single-photon mechanism is not dominant for the $J/\psi$ decays to the  $2K2\pi$ and 
$4K$ final states.
\section{\boldmath The search for the $X(3872)$}
\begin{figure}[tbh]
\includegraphics[width=0.9\linewidth]{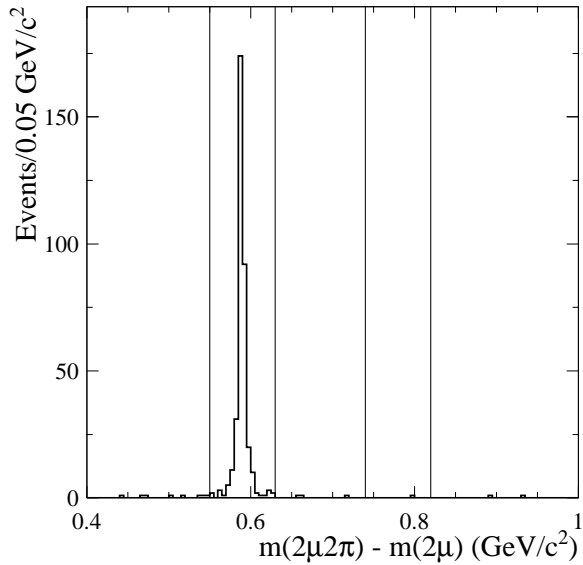}
\caption{
The $m(2\mu2\pi)-m(2\mu)$ difference for events with $m(2\mu)$ near 
the $J/\psi$ mass and both particles identified as muons. The lines show
the $\psi(2S)$ and $X(3872)$ selection regions.
}
\label{m2mu2pi2}
\end{figure}
Events in the $J/\psi$ mass region can be used to search for the new narrow
$X(3872)$ state reported by Belle~\cite{BelleX}.  If this state has $J^{PC} =
1^{--}$ it could be seen in ISR production via the decay to
$J/\psi\pi^+\pi^-$, just as for the $\psi(2S)$ (see Fig.~\ref{jpsi}).
We acknowledge that
such a $J^{PC}$ assignment is improbable since it would permit
the unobserved decay to $D\Dbar$, which would make the $X(3872)$ much broader.
If the $J/\psi$ is selected in its decay to two muons, this state can be
searched for as a peak at 3872~\mevcc in the four-charged-track
sample. 
Using our ISR data, we can set an upper limit on the
product $B_{X\to J/\psi\pipi}\cdot\Gamma^{X}_{ee}$.  To improve
background rejection, we require positive muon identification on two tracks,
in addition to having the invariant mass of the two muons fall in the
interval [3.05--3.15]~\gevcc. Figure~\ref{m2mu2pi2} displays the difference
between the mass of the four-track final state and that of the two
muons. The vertical lines show the selected regions for those events from
$\psi(2S)$ decay and for the $X(3872)$ search. The number of events within the
$\psi(2S)$ interval is used to estimate the di-muon identification efficiency,
which is found to be 0.61 for this study.  Events in the interval [0.65--1.0]
~\gevcc are used to estimate the background within the expected region for
$X(3872)$ decay, namely [0.74--0.82]~\gevcc.  One event is found within this
interval where 1.4 are expected from background; this yields an upper limit
of 3.0 events at the 90\% confidence level.  From this limit an upper limit on
the product of the branching fraction of the $X(3872)$ to $J/\psi\pi^+\pi^-$
and the \epem width of $X(3872)$ is obtained through the chain
\begin{eqnarray}
 B_{X\to J/\psi\pi^+\pi^-} \cdot B_{J/\psi\to 2\mu}\cdot\Gamma^{X}_{ee} &=&
 \nonumber \\ 
 \frac{N(X\to J/\psi\pi^+\pi^-)\cdot m_{X}^2}
      {6\cdot\pi^2\cdot d{\cal L}/dE\cdot\epsilon_{\rm MC}\cdot
  C\cdot 0.61} &<& 0.37~\ev\ ,
\end{eqnarray}
where $d{\cal L}/dE = 34.7$ nb$^{-1}$/\mev is the ISR luminosity at the $X(3872)$
mass, and $\epsilon_{\rm MC} = 0.25\pm 0.01$ is the acceptance from simulation.

Using the value $B_{J/\psi\to 2\mu} = 0.0588\pm 0.0010$~\cite{PDG}, we
extract the upper limit
$$B_{X(3872)\to J/\psi\pi^+\pi^-}\cdot\Gamma^{X}_{ee} < 6.2~\ev
  \ \mbox{at 90\% C.L.}$$

This result is the best upper limit to date and can be compared with the
result of a similar study performed by BES~\cite{BESX}.

\section{Summary}
\label{sec:Summary}
\noindent
The photon energy and charged particle momentum 
resolutions together with the particle
identification capabilities of the \babar\ detector permit the
reconstruction of
$\pipi\pipi$, $K^+K^-\pipi$ 
and $K^+K^-K^+K^-$ final states produced at low  effective c.m.\@ energy 
via ISR in data taken in the $\Upsilon(4S)$  mass region.

The analysis shows that luminosity and efficiency can be understood with
2--4\% accuracy, and that ISR production yields      
useful measurements of $R$, the ratio of the hadronic to di-muon cross 
section values, in the low-energy regime of \epem collisions.

The selected multi-hadronic final states in the broad range of accessible
energy provide new 
information on hadron spectroscopy. The  
observed $\ep\en\to\pipi\pipi$ cross section provides evidence of
resonant structure, with preferred       
quasi-two-body production of $a{1} (1260)\pi$. For the first time
there is  an indication of a
$f_{0}(1370)\rho(770)$ contribution to the final state. 
However a detailed understanding of the four-pion final state requires
additional information  from states such as $\pipi\pi^0\pi^0$.

The cross section measurements for the reaction $\epem\to K^+
K^-\pipi$ present a significant improvement upon existing data with about 15\% 
systematic uncertainty. In
addition, the final state exhibits complex resonance
sub-structure. Clear signals for $K^{*0}(892)$, $\phi$, and $\rho$ are
observed; there is evidence of $K^{*}_{2}(1430)$ production; and there
are low-mass enhancements in the $K^{*0}(892)K^\pm$, $K^{*0}(892)\pi$,
$\phi\pi$, and $K \rho$ subsystems. It is difficult to
disentangle these contributions to the final state, and we make no
attempt to do so in this paper.

The energy dependence of the cross section for the reaction $\epem\to K^+
K^- K^+ K^-$ from threshold to 4.5~\gev has been measured for the first
time with about 20\% systematic uncertainty. In contrast to the $K^+ K^-\pipi$ 
final state, the four-kaon
state shows no clear mass structure in the two- and three-body
sub-systems. The absence of a clear $\phi$ signal in the $K^+ K^-$ mass
distribution is unexpected.

The analyzed 89~\invfb of \babar\ data in the 1.4--4.5~\gevcc mass range are
already better in quality and precision than the direct measurements from
the DCI and ADONE machines, and do not suffer from the relative
normalization uncertainties which seem to exist for direct
measurements of these final states.  

The ISR events allow a study of $J/\psi$ and $\psi(2S)$ production,
and the measurement of the product of decay branching fractions and
\epem width of the $J/\psi$ with the best accuracy to date. The results
are as follows:
\begin{eqnarray*}
\lefteqn{B_{J/\psi\to 4\pi}\cdot\Gamma^{J/\psi}_{ee}} \\
  &\qquad =& (1.95\pm0.14\pm0.13)\times 10^{-2}~\kev\ ,\\
\lefteqn{B_{\psi(2S)\to J/\psi\pi^+\pi^-} \cdot B_{J/\psi\to 2\mu}
    \cdot \Gamma^{\psi(2S)}_{ee}} \\
  &\qquad =& (4.50\pm0.18\pm0.22)\times 10^{-2}~\kev\ ,\\
\lefteqn{B_{J/\psi\to 2K2\pi}\cdot\Gamma^{J/\psi}_{ee}} \\
  &\qquad =& (3.36\pm0.27\pm0.27)\times 10^{-2}~\kev\ ,\\
\lefteqn{B_{J/\psi\to 4K}\cdot\Gamma^{J/\psi}_{ee}} \\
  &\qquad =& (4.0\pm0.7\pm0.6)\times 10^{-3}~\kev\ .
\end{eqnarray*}

The dominance of the single-photon-decay  mechanism for
$J/\psi\to\pipi\pipi$ has been demonstrated by comparison with the continuum
cross section $\epem\to\pipi\pipi$. 

Under the assumption that the $X(3872)$ has $J^{PC}=1^{--}$, we have
obtained the best upper limit to date on the product of its branching
fraction to $J/\psi\pipi$ and its \epem width:
$$B_{X(3872)\to J/\psi\pi^+\pi^-}\cdot\Gamma^{X}_{ee} < 6.2~\ev
  \ \mbox{at 90\% CL.}$$

\section{Acknowledgments}
\label{sec:Acknowledgments}
The authors wish to thank Prof.\ J.\ H.\ Kuehn and H.\ Czyz for 
developing the Monte Carlo generator for ISR processes which was incorporated
into \babar\  analysis system.

We are grateful for the 
extraordinary contributions of our \pep2\ colleagues in
achieving the excellent luminosity and machine conditions
that have made this work possible.
The success of this project also relies critically on the 
expertise and dedication of the computing organizations that 
support \babar.
The collaborating institutions wish to thank 
SLAC for its support and the kind hospitality extended to them. 
This work is supported by the
US Department of Energy
and National Science Foundation, the
Natural Sciences and Engineering Research Council (Canada),
Institute of High Energy Physics (China), the
Commissariat \`a l'Energie Atomique and
Institut National de Physique Nucl\'eaire et de Physique des Particules
(France), the
Bundesministerium f\"ur Bildung und Forschung and
Deutsche Forschungsgemeinschaft
(Germany), the
Istituto Nazionale di Fisica Nucleare (Italy),
the Foundation for Fundamental Research on Matter (The Netherlands),
the Research Council of Norway, the
Ministry of Science and Technology of the Russian Federation, and the
Particle Physics and Astronomy Research Council (United Kingdom). 
Individuals have received support from 
CONACyT (Mexico),
the A. P. Sloan Foundation, 
the Research Corporation,
and the Alexander von Humboldt Foundation.

\begin{table*}
\caption{Summary of $\ep\en\to\pi^+\pi^-\pi^+\pi^-$ 
cross section measurement.
"Dressed" and "Undressed" (without vacuum polarization) cross sections are 
presented. Errors are statistical only.}
\label{4pi_tab}
\begin{ruledtabular}
\begin{tabular}{ c c c c c c c c c }
$E_{\rm c.m.}$ (GeV) & $\sigma$ (nb) &  $\sigma_{\rm noVP}$ (nb) 
& $E_{\rm c.m.}$ (GeV) & $\sigma$ (nb) &  $\sigma_{\rm noVP}$ (nb) 
& $E_{\rm c.m.}$ (GeV) & $\sigma$ (nb) &  $\sigma_{\rm noVP}$ (nb) 
\\
\hline
 0.6125 &  0.00 $\pm$  0.19 &  0.00 $\pm$  0.19 & 1.8125 & 10.06 $\pm$  0.40 &  9.68 $\pm$  0.38 & 3.0125 &  0.61 $\pm$  0.10 &  0.60 $\pm$  0.10 \\ 
 0.6375 &  0.22 $\pm$  0.13 &  0.22 $\pm$  0.13 & 1.8375 &  8.29 $\pm$  0.37 &  7.98 $\pm$  0.36 & 3.0375 &  0.85 $\pm$  0.10 &  0.84 $\pm$  0.10 \\ 
 0.6625 &  0.07 $\pm$  0.16 &  0.07 $\pm$  0.16 & 1.8625 &  6.99 $\pm$  0.33 &  6.73 $\pm$  0.32 & 3.0625 &  0.58 $\pm$  0.09 &  0.59 $\pm$  0.09 \\ 
 0.6875 &  0.09 $\pm$  0.07 &  0.09 $\pm$  0.07 & 1.8875 &  6.86 $\pm$  0.33 &  6.60 $\pm$  0.32 & 3.0875 &  1.53 $\pm$  0.13 &  0.43 $\pm$  0.04 \\ 
 0.7125 &  0.00 $\pm$  0.00 &  0.00 $\pm$  0.00 & 1.9125 &  6.23 $\pm$  0.32 &  6.00 $\pm$  0.31 & 3.1125 &  1.02 $\pm$  0.12 &  0.78 $\pm$  0.09 \\ 
 0.7375 &  0.07 $\pm$  0.05 &  0.07 $\pm$  0.05 & 1.9375 &  6.55 $\pm$  0.31 &  6.31 $\pm$  0.30 & 3.1375 &  0.31 $\pm$  0.08 &  0.28 $\pm$  0.07 \\ 
 0.7625 &  0.03 $\pm$  0.03 &  0.03 $\pm$  0.03 & 1.9625 &  6.29 $\pm$  0.31 &  6.06 $\pm$  0.30 & 3.1625 &  0.44 $\pm$  0.09 &  0.41 $\pm$  0.08 \\ 
 0.7875 &  0.06 $\pm$  0.08 &  0.06 $\pm$  0.08 & 1.9875 &  5.92 $\pm$  0.31 &  5.70 $\pm$  0.30 & 3.1875 &  0.44 $\pm$  0.08 &  0.41 $\pm$  0.07 \\ 
 0.8125 &  0.08 $\pm$  0.05 &  0.08 $\pm$  0.05 & 2.0125 &  5.48 $\pm$  0.30 &  5.28 $\pm$  0.29 & 3.2125 &  0.51 $\pm$  0.08 &  0.48 $\pm$  0.08 \\ 
 0.8375 &  0.11 $\pm$  0.08 &  0.10 $\pm$  0.08 & 2.0375 &  5.72 $\pm$  0.29 &  5.51 $\pm$  0.28 & 3.2375 &  0.23 $\pm$  0.07 &  0.22 $\pm$  0.07 \\ 
 0.8625 &  0.28 $\pm$  0.09 &  0.27 $\pm$  0.09 & 2.0625 &  5.38 $\pm$  0.28 &  5.18 $\pm$  0.27 & 3.2625 &  0.29 $\pm$  0.07 &  0.27 $\pm$  0.07 \\ 
 0.8875 &  0.37 $\pm$  0.09 &  0.35 $\pm$  0.09 & 2.0875 &  5.50 $\pm$  0.28 &  5.29 $\pm$  0.27 & 3.2875 &  0.29 $\pm$  0.08 &  0.27 $\pm$  0.08 \\ 
 0.9125 &  0.44 $\pm$  0.11 &  0.42 $\pm$  0.11 & 2.1125 &  4.60 $\pm$  0.26 &  4.43 $\pm$  0.25 & 3.3125 &  0.37 $\pm$  0.08 &  0.35 $\pm$  0.08 \\ 
 0.9375 &  0.36 $\pm$  0.11 &  0.35 $\pm$  0.11 & 2.1375 &  4.78 $\pm$  0.26 &  4.60 $\pm$  0.25 & 3.3375 &  0.35 $\pm$  0.08 &  0.33 $\pm$  0.08 \\ 
 0.9625 &  0.78 $\pm$  0.14 &  0.76 $\pm$  0.14 & 2.1625 &  4.73 $\pm$  0.26 &  4.55 $\pm$  0.25 & 3.3625 &  0.31 $\pm$  0.07 &  0.29 $\pm$  0.07 \\ 
 0.9875 &  0.94 $\pm$  0.16 &  0.92 $\pm$  0.16 & 2.1875 &  3.82 $\pm$  0.24 &  3.68 $\pm$  0.23 & 3.3875 &  0.31 $\pm$  0.07 &  0.29 $\pm$  0.07 \\ 
 1.0125 &  1.14 $\pm$  0.17 &  1.13 $\pm$  0.17 & 2.2125 &  3.49 $\pm$  0.24 &  3.36 $\pm$  0.23 & 3.4125 &  0.31 $\pm$  0.07 &  0.30 $\pm$  0.07 \\ 
 1.0375 &  1.76 $\pm$  0.20 &  1.64 $\pm$  0.19 & 2.2375 &  3.55 $\pm$  0.23 &  3.42 $\pm$  0.22 & 3.4375 &  0.09 $\pm$  0.07 &  0.09 $\pm$  0.07 \\ 
 1.0625 &  2.65 $\pm$  0.22 &  2.52 $\pm$  0.21 & 2.2625 &  3.43 $\pm$  0.23 &  3.30 $\pm$  0.22 & 3.4625 &  0.23 $\pm$  0.07 &  0.22 $\pm$  0.07 \\ 
 1.0875 &  3.07 $\pm$  0.25 &  2.93 $\pm$  0.24 & 2.2875 &  3.11 $\pm$  0.23 &  2.99 $\pm$  0.22 & 3.4875 &  0.31 $\pm$  0.06 &  0.30 $\pm$  0.06 \\ 
 1.1125 &  3.82 $\pm$  0.28 &  3.66 $\pm$  0.27 & 2.3125 &  2.69 $\pm$  0.20 &  2.59 $\pm$  0.19 & 3.5125 &  0.26 $\pm$  0.06 &  0.25 $\pm$  0.06 \\ 
 1.1375 &  5.02 $\pm$  0.33 &  4.82 $\pm$  0.32 & 2.3375 &  3.13 $\pm$  0.21 &  3.01 $\pm$  0.20 & 3.5375 &  0.14 $\pm$  0.06 &  0.13 $\pm$  0.06 \\ 
 1.1625 &  7.10 $\pm$  0.37 &  6.83 $\pm$  0.36 & 2.3625 &  2.51 $\pm$  0.19 &  2.42 $\pm$  0.18 & 3.5625 &  0.13 $\pm$  0.06 &  0.12 $\pm$  0.06 \\ 
 1.1875 &  7.97 $\pm$  0.39 &  7.67 $\pm$  0.38 & 2.3875 &  2.11 $\pm$  0.19 &  2.03 $\pm$  0.18 & 3.5875 &  0.12 $\pm$  0.06 &  0.12 $\pm$  0.06 \\ 
 1.2125 & 10.56 $\pm$  0.45 & 10.17 $\pm$  0.43 & 2.4125 &  2.30 $\pm$  0.18 &  2.21 $\pm$  0.17 & 3.6125 &  0.12 $\pm$  0.05 &  0.12 $\pm$  0.05 \\
 1.2375 & 12.30 $\pm$  0.47 & 11.86 $\pm$  0.45 & 2.4375 &  1.94 $\pm$  0.18 &  1.87 $\pm$  0.17 & 3.6375 &  0.09 $\pm$  0.06 &  0.09 $\pm$  0.06 \\
 1.2625 & 13.48 $\pm$  0.51 & 13.00 $\pm$  0.49 & 2.4625 &  2.18 $\pm$  0.16 &  2.10 $\pm$  0.15 & 3.6625 &  0.00 $\pm$  0.06 &  0.00 $\pm$  0.06 \\
 1.2875 & 16.02 $\pm$  0.53 & 15.46 $\pm$  0.51 & 2.4875 &  1.76 $\pm$  0.17 &  1.69 $\pm$  0.16 & 3.6875 &  0.12 $\pm$  0.05 &  0.10 $\pm$  0.04 \\
 1.3125 & 18.27 $\pm$  0.57 & 17.64 $\pm$  0.55 & 2.5125 &  1.73 $\pm$  0.16 &  1.67 $\pm$  0.15 & 3.7125 &  0.06 $\pm$  0.05 &  0.06 $\pm$  0.05 \\
 1.3375 & 20.27 $\pm$  0.60 & 19.57 $\pm$  0.58 & 2.5375 &  1.62 $\pm$  0.17 &  1.56 $\pm$  0.16 & 3.7375 &  0.11 $\pm$  0.05 &  0.10 $\pm$  0.05 \\
 1.3625 & 21.70 $\pm$  0.61 & 20.97 $\pm$  0.59 & 2.5625 &  1.69 $\pm$  0.15 &  1.63 $\pm$  0.14 & 3.7625 &  0.22 $\pm$  0.06 &  0.21 $\pm$  0.06 \\
 1.3875 & 24.90 $\pm$  0.66 & 24.06 $\pm$  0.64 & 2.5875 &  1.49 $\pm$  0.15 &  1.43 $\pm$  0.14 & 3.7875 &  0.08 $\pm$  0.05 &  0.08 $\pm$  0.05 \\
 1.4125 & 27.05 $\pm$  0.67 & 26.16 $\pm$  0.65 & 2.6125 &  1.50 $\pm$  0.15 &  1.44 $\pm$  0.14 & 3.8125 &  0.12 $\pm$  0.05 &  0.11 $\pm$  0.05 \\
 1.4375 & 28.33 $\pm$  0.68 & 27.37 $\pm$  0.66 & 2.6375 &  1.33 $\pm$  0.15 &  1.28 $\pm$  0.14 & 3.8375 &  0.04 $\pm$  0.05 &  0.04 $\pm$  0.05 \\
 1.4625 & 29.32 $\pm$  0.70 & 28.33 $\pm$  0.68 & 2.6625 &  1.24 $\pm$  0.13 &  1.19 $\pm$  0.12 & 3.8625 &  0.10 $\pm$  0.05 &  0.09 $\pm$  0.05 \\
 1.4875 & 30.20 $\pm$  0.72 & 29.18 $\pm$  0.70 & 2.6875 &  1.07 $\pm$  0.13 &  1.03 $\pm$  0.12 & 3.8875 &  0.13 $\pm$  0.05 &  0.12 $\pm$  0.05 \\
 1.5125 & 29.82 $\pm$  0.70 & 28.78 $\pm$  0.68 & 2.7125 &  1.01 $\pm$  0.13 &  0.97 $\pm$  0.12 & 3.9125 &  0.17 $\pm$  0.04 &  0.16 $\pm$  0.04 \\
 1.5375 & 28.77 $\pm$  0.67 & 27.77 $\pm$  0.65 & 2.7375 &  0.94 $\pm$  0.13 &  0.90 $\pm$  0.13 & 3.9375 &  0.08 $\pm$  0.05 &  0.08 $\pm$  0.05 \\
 1.5625 & 26.43 $\pm$  0.65 & 25.51 $\pm$  0.63 & 2.7625 &  1.01 $\pm$  0.12 &  0.97 $\pm$  0.12 & 3.9625 &  0.06 $\pm$  0.05 &  0.06 $\pm$  0.05 \\
 1.5875 & 26.03 $\pm$  0.63 & 25.13 $\pm$  0.61 & 2.7875 &  1.05 $\pm$  0.12 &  1.01 $\pm$  0.12 & 3.9875 &  0.07 $\pm$  0.05 &  0.07 $\pm$  0.05 \\
 1.6125 & 22.93 $\pm$  0.61 & 22.13 $\pm$  0.59 & 2.8125 &  0.88 $\pm$  0.11 &  0.85 $\pm$  0.11 & 4.0125 &  0.05 $\pm$  0.05 &  0.05 $\pm$  0.05 \\
 1.6375 & 22.05 $\pm$  0.59 & 21.28 $\pm$  0.57 & 2.8375 &  0.86 $\pm$  0.12 &  0.83 $\pm$  0.12 & 4.0375 &  0.07 $\pm$  0.04 &  0.07 $\pm$  0.04 \\
 1.6625 & 19.84 $\pm$  0.57 & 19.13 $\pm$  0.55 & 2.8625 &  0.75 $\pm$  0.11 &  0.72 $\pm$  0.11 & 4.0625 &  0.08 $\pm$  0.04 &  0.08 $\pm$  0.04 \\
 1.6875 & 17.79 $\pm$  0.52 & 17.16 $\pm$  0.50 & 2.8875 &  0.77 $\pm$  0.11 &  0.74 $\pm$  0.11 & 4.0875 &  0.08 $\pm$  0.04 &  0.08 $\pm$  0.04 \\
 1.7125 & 16.22 $\pm$  0.50 & 15.63 $\pm$  0.48 & 2.9125 &  0.81 $\pm$  0.12 &  0.78 $\pm$  0.12 & 4.1500 &  0.03 $\pm$  0.02 &  0.03 $\pm$  0.02 \\
 1.7375 & 14.98 $\pm$  0.48 & 14.43 $\pm$  0.46 & 2.9375 &  0.51 $\pm$  0.11 &  0.49 $\pm$  0.11 & 4.2500 &  0.01 $\pm$  0.02 &  0.01 $\pm$  0.02 \\
 1.7625 & 12.92 $\pm$  0.45 & 12.44 $\pm$  0.43 & 2.9625 &  0.59 $\pm$  0.10 &  0.57 $\pm$  0.10 & 4.3500 &  0.02 $\pm$  0.02 &  0.02 $\pm$  0.02 \\
 1.7875 & 10.75 $\pm$  0.42 & 10.35 $\pm$  0.40 & 2.9875 &  0.64 $\pm$  0.10 &  0.62 $\pm$  0.10 & 4.4500 &  0.00 $\pm$  0.02 &  0.00 $\pm$  0.02 \\
\end{tabular}
\end{ruledtabular}
\end{table*}

\begin{table*}
\caption{Summary of $\ep\en\to K^+ K^-\pi^+\pi^-$                                                                  
cross section measurement.                                                                                         
"Dressed" and "Undressed" (without vacuum polarization) cross sections are                                         
presented. Errors are statistical only.}                                                                           
\label{2k2pi_tab} 
\begin{ruledtabular}
\begin{tabular}{ c c c c c c c c c }
$E_{\rm c.m.}$ (GeV) & $\sigma$ (nb) &  $\sigma_{\rm noVP}$ (nb)
& $E_{\rm c.m.}$ (GeV) & $\sigma$ (nb) &  $\sigma_{\rm noVP}$ (nb)
& $E_{\rm c.m.}$ (GeV) & $\sigma$ (nb) &  $\sigma_{\rm noVP}$ (nb)
\\
\hline
 1.5125 &  0.00 $\pm$  0.00 &  0.00 $\pm$  0.00 & 2.5125 &  1.27 $\pm$  0.19 &  1.22 $\pm$  0.18 & 3.5125 &  0.33 $\pm$  0.07 &  0.32 $\pm$  0.07 \\ 
 1.5375 &  0.03 $\pm$  0.05 &  0.03 $\pm$  0.05 & 2.5375 &  1.20 $\pm$  0.18 &  1.15 $\pm$  0.17 & 3.5375 &  0.29 $\pm$  0.07 &  0.28 $\pm$  0.07 \\ 
 1.5625 &  0.11 $\pm$  0.05 &  0.11 $\pm$  0.05 & 2.5625 &  1.58 $\pm$  0.19 &  1.52 $\pm$  0.18 & 3.5625 &  0.22 $\pm$  0.06 &  0.21 $\pm$  0.06 \\ 
 1.5875 &  0.29 $\pm$  0.08 &  0.28 $\pm$  0.08 & 2.5875 &  1.22 $\pm$  0.17 &  1.17 $\pm$  0.16 & 3.5875 &  0.07 $\pm$  0.06 &  0.07 $\pm$  0.06 \\ 
 1.6125 &  0.58 $\pm$  0.13 &  0.56 $\pm$  0.13 & 2.6125 &  1.02 $\pm$  0.16 &  0.98 $\pm$  0.15 & 3.6125 &  0.08 $\pm$  0.05 &  0.08 $\pm$  0.05 \\ 
 1.6375 &  0.96 $\pm$  0.15 &  0.93 $\pm$  0.14 & 2.6375 &  1.03 $\pm$  0.16 &  0.99 $\pm$  0.15 & 3.6375 &  0.08 $\pm$  0.05 &  0.08 $\pm$  0.05 \\ 
 1.6625 &  1.22 $\pm$  0.19 &  1.18 $\pm$  0.18 & 2.6625 &  0.83 $\pm$  0.17 &  0.80 $\pm$  0.16 & 3.6625 &  0.16 $\pm$  0.06 &  0.16 $\pm$  0.06 \\ 
 1.6875 &  1.60 $\pm$  0.22 &  1.54 $\pm$  0.21 & 2.6875 &  0.69 $\pm$  0.14 &  0.66 $\pm$  0.13 & 3.6875 &  0.21 $\pm$  0.07 &  0.17 $\pm$  0.06 \\ 
 1.7125 &  1.78 $\pm$  0.25 &  1.71 $\pm$  0.24 & 2.7125 &  1.18 $\pm$  0.17 &  1.13 $\pm$  0.16 & 3.7125 &  0.24 $\pm$  0.07 &  0.22 $\pm$  0.06 \\ 
 1.7375 &  2.67 $\pm$  0.30 &  2.57 $\pm$  0.29 & 2.7375 &  0.67 $\pm$  0.15 &  0.64 $\pm$  0.14 & 3.7375 &  0.11 $\pm$  0.06 &  0.10 $\pm$  0.06 \\ 
 1.7625 &  2.81 $\pm$  0.33 &  2.70 $\pm$  0.32 & 2.7625 &  0.98 $\pm$  0.14 &  0.94 $\pm$  0.13 & 3.7625 &  0.23 $\pm$  0.06 &  0.22 $\pm$  0.06 \\ 
 1.7875 &  2.99 $\pm$  0.34 &  2.88 $\pm$  0.33 & 2.7875 &  0.78 $\pm$  0.14 &  0.75 $\pm$  0.13 & 3.7875 &  0.13 $\pm$  0.05 &  0.12 $\pm$  0.05 \\ 
 1.8125 &  3.72 $\pm$  0.36 &  3.58 $\pm$  0.35 & 2.8125 &  0.72 $\pm$  0.13 &  0.69 $\pm$  0.13 & 3.8125 &  0.06 $\pm$  0.05 &  0.06 $\pm$  0.05 \\ 
 1.8375 &  4.16 $\pm$  0.40 &  4.00 $\pm$  0.38 & 2.8375 &  0.77 $\pm$  0.13 &  0.74 $\pm$  0.13 & 3.8375 &  0.13 $\pm$  0.06 &  0.12 $\pm$  0.06 \\ 
 1.8625 &  4.58 $\pm$  0.41 &  4.41 $\pm$  0.39 & 2.8625 &  0.63 $\pm$  0.13 &  0.61 $\pm$  0.13 & 3.8625 &  0.08 $\pm$  0.06 &  0.08 $\pm$  0.06 \\ 
 1.8875 &  4.34 $\pm$  0.41 &  4.18 $\pm$  0.39 & 2.8875 &  0.28 $\pm$  0.12 &  0.27 $\pm$  0.12 & 3.8875 &  0.08 $\pm$  0.05 &  0.08 $\pm$  0.05 \\ 
 1.9125 &  4.18 $\pm$  0.38 &  4.03 $\pm$  0.37 & 2.9125 &  0.48 $\pm$  0.13 &  0.46 $\pm$  0.13 & 3.9125 &  0.07 $\pm$  0.04 &  0.07 $\pm$  0.04 \\ 
 1.9375 &  4.70 $\pm$  0.42 &  4.53 $\pm$  0.40 & 2.9375 &  0.53 $\pm$  0.11 &  0.51 $\pm$  0.11 & 3.9375 &  0.07 $\pm$  0.05 &  0.07 $\pm$  0.05 \\ 
 1.9625 &  4.01 $\pm$  0.38 &  3.86 $\pm$  0.37 & 2.9625 &  0.44 $\pm$  0.12 &  0.43 $\pm$  0.12 & 3.9625 &  0.10 $\pm$  0.05 &  0.10 $\pm$  0.05 \\ 
 1.9875 &  4.11 $\pm$  0.38 &  3.96 $\pm$  0.37 & 2.9875 &  0.54 $\pm$  0.11 &  0.53 $\pm$  0.11 & 3.9875 &  0.03 $\pm$  0.05 &  0.03 $\pm$  0.05 \\ 
 2.0125 &  3.08 $\pm$  0.33 &  2.97 $\pm$  0.32 & 3.0125 &  0.48 $\pm$  0.12 &  0.47 $\pm$  0.12 & 4.0125 &  0.04 $\pm$  0.04 &  0.04 $\pm$  0.04 \\
 2.0375 &  3.19 $\pm$  0.33 &  3.07 $\pm$  0.32 & 3.0375 &  0.39 $\pm$  0.11 &  0.39 $\pm$  0.11 & 4.0375 &  0.13 $\pm$  0.05 &  0.12 $\pm$  0.05 \\
 2.0625 &  3.32 $\pm$  0.35 &  3.20 $\pm$  0.34 & 3.0625 &  0.46 $\pm$  0.11 &  0.47 $\pm$  0.11 & 4.0625 &  0.00 $\pm$  0.00 &  0.00 $\pm$  0.00 \\
 2.0875 &  2.79 $\pm$  0.32 &  2.69 $\pm$  0.31 & 3.0875 &  2.40 $\pm$  0.24 &  0.67 $\pm$  0.07 & 4.0875 &  0.02 $\pm$  0.04 &  0.02 $\pm$  0.04 \\
 2.1125 &  3.31 $\pm$  0.32 &  3.19 $\pm$  0.31 & 3.1125 &  1.30 $\pm$  0.17 &  1.00 $\pm$  0.13 & 4.1125 &  0.05 $\pm$  0.04 &  0.05 $\pm$  0.04 \\
 2.1375 &  2.84 $\pm$  0.32 &  2.73 $\pm$  0.31 & 3.1375 &  0.59 $\pm$  0.11 &  0.53 $\pm$  0.10 & 4.1375 &  0.00 $\pm$  0.04 &  0.00 $\pm$  0.04 \\
 2.1625 &  2.48 $\pm$  0.28 &  2.39 $\pm$  0.27 & 3.1625 &  0.48 $\pm$  0.11 &  0.44 $\pm$  0.10 & 4.1625 &  0.04 $\pm$  0.04 &  0.04 $\pm$  0.04 \\
 2.1875 &  2.83 $\pm$  0.30 &  2.72 $\pm$  0.29 & 3.1875 &  0.32 $\pm$  0.10 &  0.30 $\pm$  0.09 & 4.1875 &  0.04 $\pm$  0.05 &  0.04 $\pm$  0.05 \\
 2.2125 &  2.00 $\pm$  0.25 &  1.92 $\pm$  0.24 & 3.2125 &  0.57 $\pm$  0.11 &  0.54 $\pm$  0.10 & 4.2125 &  0.06 $\pm$  0.04 &  0.06 $\pm$  0.04 \\
 2.2375 &  1.79 $\pm$  0.24 &  1.72 $\pm$  0.23 & 3.2375 &  0.46 $\pm$  0.10 &  0.43 $\pm$  0.09 & 4.2375 &  0.01 $\pm$  0.04 &  0.01 $\pm$  0.04 \\
 2.2625 &  2.18 $\pm$  0.26 &  2.10 $\pm$  0.25 & 3.2625 &  0.40 $\pm$  0.09 &  0.38 $\pm$  0.08 & 4.2625 &  0.13 $\pm$  0.04 &  0.12 $\pm$  0.04 \\
 2.2875 &  1.53 $\pm$  0.23 &  1.47 $\pm$  0.22 & 3.2875 &  0.32 $\pm$  0.08 &  0.30 $\pm$  0.08 & 4.2875 &  0.04 $\pm$  0.04 &  0.04 $\pm$  0.04 \\
 2.3125 &  1.91 $\pm$  0.24 &  1.84 $\pm$  0.23 & 3.3125 &  0.40 $\pm$  0.09 &  0.38 $\pm$  0.09 & 4.3125 &  0.02 $\pm$  0.04 &  0.02 $\pm$  0.04 \\
 2.3375 &  1.73 $\pm$  0.24 &  1.67 $\pm$  0.23 & 3.3375 &  0.25 $\pm$  0.07 &  0.24 $\pm$  0.07 & 4.3375 &  0.06 $\pm$  0.03 &  0.06 $\pm$  0.03 \\
 2.3625 &  1.77 $\pm$  0.21 &  1.70 $\pm$  0.20 & 3.3625 &  0.40 $\pm$  0.09 &  0.38 $\pm$  0.09 & 4.3625 &  0.00 $\pm$  0.04 &  0.00 $\pm$  0.04 \\
 2.3875 &  1.54 $\pm$  0.22 &  1.48 $\pm$  0.21 & 3.3875 &  0.24 $\pm$  0.07 &  0.23 $\pm$  0.07 & 4.3875 &  0.02 $\pm$  0.04 &  0.02 $\pm$  0.04 \\
 2.4125 &  2.01 $\pm$  0.23 &  1.93 $\pm$  0.22 & 3.4125 &  0.22 $\pm$  0.08 &  0.21 $\pm$  0.08 & 4.4125 &  0.05 $\pm$  0.04 &  0.05 $\pm$  0.04 \\
 2.4375 &  1.38 $\pm$  0.20 &  1.33 $\pm$  0.19 & 3.4375 &  0.00 $\pm$  0.00 &  0.00 $\pm$  0.00 & 4.4375 &  0.05 $\pm$  0.03 &  0.05 $\pm$  0.03 \\
 2.4625 &  1.61 $\pm$  0.22 &  1.55 $\pm$  0.21 & 3.4625 &  0.24 $\pm$  0.08 &  0.23 $\pm$  0.08 & 4.4625 &  0.05 $\pm$  0.03 &  0.05 $\pm$  0.03 \\
 2.4875 &  1.84 $\pm$  0.21 &  1.77 $\pm$  0.20 & 3.4875 &  0.21 $\pm$  0.07 &  0.20 $\pm$  0.07 & 4.4875 &  0.06 $\pm$  0.04 &  0.06 $\pm$  0.04 \\
\end{tabular}                                                                                                      
\end{ruledtabular}                                                                                                 
\end{table*}

\begin{table*}                                                                                                      
\caption{Summary of $\ep\en\to K^+ K^- K^+ K^-$                                                                    
cross section measurement.                                                                                         
"Dressed" and "Undressed" (without vacuum polarization) cross sections are                                         
presented. Errors are statistical only.}
\label{4k_tab}                                                                                                    
\begin{ruledtabular}                                                                                               
\begin{tabular}{ c c c c c c c c c }                                                                               
$E_{\rm c.m.}$ (GeV) & $\sigma$ (nb) &  $\sigma_{\rm noVP}$ (nb)                                                   
& $E_{\rm c.m.}$ (GeV) & $\sigma$ (nb) &  $\sigma_{\rm noVP}$ (nb)                                                 
& $E_{\rm c.m.}$ (GeV) & $\sigma$ (nb) &  $\sigma_{\rm noVP}$ (nb)                                                 
\\                                                                                                                 
\hline
 2.0312 &  0.00 $\pm$  0.01 &  0.00 $\pm$  0.00 & 2.9062 &  0.07 $\pm$  0.02 &  0.06 $\pm$  0.02 &  3.7188 &  0.00 $\pm$  0.01 &  0.00 $\pm$  0.01 \\
 2.0938 &  0.02 $\pm$  0.01 &  0.02 $\pm$  0.01 & 2.9688 &  0.08 $\pm$  0.02 &  0.08 $\pm$  0.02 &  3.7812 &  0.05 $\pm$  0.01 &  0.04 $\pm$  0.01 \\
 2.1562 &  0.05 $\pm$  0.01 &  0.05 $\pm$  0.01 & 3.0312 &  0.06 $\pm$  0.02 &  0.06 $\pm$  0.02 &  3.8438 &  0.03 $\pm$  0.01 &  0.03 $\pm$  0.01 \\
 2.2188 &  0.08 $\pm$  0.02 &  0.08 $\pm$  0.02 & 3.0938 &  0.19 $\pm$  0.03 &  0.18 $\pm$  0.03 &  3.9062 &  0.01 $\pm$  0.01 &  0.01 $\pm$  0.01 \\
 2.2812 &  0.07 $\pm$  0.02 &  0.07 $\pm$  0.02 & 3.1562 &  0.03 $\pm$  0.02 &  0.03 $\pm$  0.02 &  3.9688 &  0.01 $\pm$  0.01 &  0.01 $\pm$  0.01 \\
 2.3438 &  0.08 $\pm$  0.02 &  0.07 $\pm$  0.02 & 3.2188 &  0.05 $\pm$  0.01 &  0.05 $\pm$  0.01 &  4.0312 &  0.03 $\pm$  0.01 &  0.03 $\pm$  0.01 \\
 2.4062 &  0.07 $\pm$  0.02 &  0.07 $\pm$  0.02 & 3.2812 &  0.03 $\pm$  0.01 &  0.03 $\pm$  0.01 &  4.0938 &  0.01 $\pm$  0.01 &  0.01 $\pm$  0.01 \\
 2.4688 &  0.08 $\pm$  0.03 &  0.08 $\pm$  0.03 & 3.3438 &  0.04 $\pm$  0.02 &  0.04 $\pm$  0.02 &  4.1562 &  0.02 $\pm$  0.01 &  0.02 $\pm$  0.01 \\
 2.5312 &  0.05 $\pm$  0.02 &  0.05 $\pm$  0.02 & 3.4062 &  0.05 $\pm$  0.02 &  0.05 $\pm$  0.02 &  4.2188 &  0.04 $\pm$  0.01 &  0.04 $\pm$  0.01 \\
 2.5938 &  0.07 $\pm$  0.02 &  0.07 $\pm$  0.02 & 3.4688 &  0.03 $\pm$  0.01 &  0.03 $\pm$  0.01 &  4.2812 &  0.00 $\pm$  0.01 &  0.00 $\pm$  0.01 \\
 2.6562 &  0.10 $\pm$  0.03 &  0.10 $\pm$  0.02 & 3.5312 &  0.05 $\pm$  0.01 &  0.05 $\pm$  0.01 &  4.3438 &  0.01 $\pm$  0.01 &  0.00 $\pm$  0.01 \\
 2.7188 &  0.13 $\pm$  0.03 &  0.12 $\pm$  0.03 & 3.5938 &  0.02 $\pm$  0.01 &  0.02 $\pm$  0.01 &  4.4062 &  0.00 $\pm$  0.01 &  0.00 $\pm$  0.01 \\
 2.7812 &  0.08 $\pm$  0.02 &  0.07 $\pm$  0.02 & 3.6562 &  0.02 $\pm$  0.01 &  0.02 $\pm$  0.01 &  4.4688 &  0.00 $\pm$  0.01 &  0.00 $\pm$  0.01 \\
 2.8438 &  0.10 $\pm$  0.02 &  0.09 $\pm$  0.02 &  & & & & & \\ 
\end{tabular}
\end{ruledtabular}
\end{table*}

\end{document}